\documentclass[twocolumn,prd,amsmath,amsfonts,nofootinbib,showpacs]{revtex4}

\newcommand{\fd}[2]{\frac{\delta #1}{\delta #2}}
\newcommand{\ud}{\text{d}}
\newcommand{\half}{{\tfrac{1}{2}}}
\newcommand{\bea}{\begin{eqnarray}}
\newcommand{\eea}{\end{eqnarray}}
\newcommand{\gb}{\bar{g}}
\newcommand{\delb}{\bar{\nabla}}
\newcommand{\Rb}{\bar{R}}
\newcommand{\Psib}{\bar{\Psi}}
\newcommand{\bL}{\bar{\mathcal{L}}}

\usepackage{mathptmx} 

\begin{document}
\title{Bootstrapping gravity: a consistent approach to energy-momentum self-coupling}
\author{Luke M. Butcher}
\email[]{l.butcher@mrao.cam.ac.uk}
\author{Michael Hobson}
\author{Anthony Lasenby}
\affiliation{Astrophysics Group, Cavendish Laboratory, J J Thomson Avenue, Cambridge CB3 0HE, UK.}
\date{\today}

\begin{abstract}
It is generally believed that coupling the graviton (a classical Fierz-Pauli massless spin-2 field) to its own energy-momentum tensor successfully recreates the dynamics of the Einstein field equations order by order; however the validity of this idea has recently been brought into doubt \cite{Pad}. Motivated by this, we present a graviton action for which energy-momentum self-coupling is indeed consistent with the Einstein field equations. The Hilbert energy-momentum tensor for this graviton is calculated explicitly and shown to supply the correct second-order term in the field equations; in contrast, the Fierz-Pauli action fails to supply the correct term. A formalism for perturbative expansions of metric-based gravitational theories is then developed, and these techniques employed to demonstrate that our graviton action is a starting point for a straightforward energy-momentum self-coupling procedure that, order by order, generates the Einstein-Hilbert action (up to a classically irrelevant surface term). The perturbative formalism is extended to include matter and a cosmological constant, and interactions between perturbations of a free matter field and the gravitational field are studied in a vacuum background. Finally, the effect of a non-vacuum background is examined, and the graviton is found to develop a non-vanishing ``mass-term'' in the action.
\end{abstract}
\maketitle

\section{Introduction}
It is a standard view in particle physics that the non-linearity of a field theory, such as those of Yang and Mills, can be equated with the notion that the field in question carries the charge of the very interaction it mediates. This idea has been brought to bear on gravity many times, and various arguments \cite{krai,gupta,deser,Deser2,Deser&Boulware,feyngrav,Ortin} aim to derive general relativity from a linear starting point by coupling gravity to the energy and momentum of all fields, including the gravitational field itself. Despite the conventional wisdom that this self-coupling process is already well understood, Padmanabhan has uncovered a number of serious problems with the standard arguments \cite{Pad}. Although we postpone an examination of Padmanabhan's analysis to appendix \ref{pad}, it suffices to express here what is, in our view, his most pertinent observation: one cannot start with \emph{linear gravity}, the Fierz-Pauli massless spin-2 action \cite{fierz,Pad}, and generate the higher-order corrections of general relativity by coupling the gravitational field to its own Hilbert energy-momentum tensor. More succinctly: one cannot derive the Einstein equations by bootstrapping gravitons\footnote{In discussions of this nature, the word \emph{graviton} is often used as a shorthand for the classical massless spin-2 field. We follow this convention to cohere with the literature, but stress that this graviton is in no way quantum mechanical. What is actually being referred to is a \emph{gravitational wave}, a classical fluctuation in the geometry of spacetime.} to their own energy and momentum.

To clarify the content of this observation, consider a perturbative expansion of the Einstein field equations $G_{\alpha\beta}=\kappa T^\mathrm{matter}_{\alpha\beta}$  about a Minkowski background: $g_{\alpha\beta}= \eta_{\alpha\beta} + h_{\alpha\beta}$. Working to second-order in $h_{\alpha\beta}$, we obtain
\bea\label{aprox}
G^{(1)}_{\alpha\beta}=- G^{(2)}_{\alpha\beta} + \kappa T^\mathrm{matter}_{\alpha\beta},
\eea
where the numbers in parenthesis denote the powers of $h_{\alpha\beta}$ the term contains. Because $G^{(1)}_{\alpha\beta}=0$ is the equation of motion for a massless spin-2 field $h_{\alpha\beta}$, the right-hand side of (\ref{aprox}) can be interpreted as this field's source. Thus a satisfying physical picture suggests itself: the gravitational field $h_{\alpha\beta}$ is induced by the energy-momentum tensor of \emph{all} fields $T_{\alpha\beta}= T^\mathrm{matter}_{\alpha\beta} + t_{\alpha\beta}$, where $t_{\alpha\beta}$  is gravity's own energy-momentum tensor, identified as $-G^{(2)}_{\alpha\beta}/\kappa$. In actuality, however, this description cannot be formulated in a straightforward manner. Although the Fierz-Pauli action $S_\mathrm{FP}$ is typically used to prescribe the dynamics of a massless spin-2 field, its Hilbert energy-momentum tensor\footnote{Although other definitions of the energy-momentum tensor exist (see \S\ref{EMT}) we must define $t_{\alpha\beta}$ according to the Hilbert's prescription (\ref{Hilbert1}) in order to maintain the analogy with $T^\mathrm{matter}_{\alpha\beta}$. This definition requires that $S_\mathrm{FP}$ be ``covariantized'' (represented in arbitrary coordinates using a \emph{flat} metric $\gamma_{\alpha\beta}$) and a functional derivative taken with respect to the metric. It is important to realise that even though $\gamma_{\alpha\beta}$ is flat, the arbitrary variations $\delta\gamma_{\alpha\beta}$ required to construct the functional derivative inevitably explore \emph{curved} metrics in a neighbourhood of $\gamma_{\alpha\beta}$. Thus ``covariantization'' is not really sufficient: the action must be generalised to a curved background spacetime. One of the key aims of this paper is to generalise $S_\mathrm{FP}$ to curved spacetime in such a way that energy-momentum self-coupling is consistent with general relativity.} 
\bea\label{Hilbert1}
t_{\alpha\beta} \equiv \frac{-1}{\sqrt{-\gamma}}\fd{S_\mathrm{FP}}{\gamma^{\alpha\beta}},
\eea
is \emph{not} proportional to $G^{(2)}_{\alpha\beta}$, and thus cannot be used as the source-term for the second-order field equations. As an alternative approach, one could introduce energy-momentum self-coupling at the level of the action: because $t_{\alpha\beta}$ is a function of $h_{\alpha\beta}$, adding the self-coupling term $t_{\alpha\beta}h^{\alpha\beta}$ to the Lagrangian yields a different result from adding $t_{\alpha\beta}$ directly to the equations of motion. Unfortunately, this procedure also fails to generate $-G^{(2)}_{\alpha\beta}/\kappa$ in the field equations.

Padmanabhan claims that these realizations bring to light a previously neglected object $S^{\alpha\beta}$ (see appendix \ref{pad}) which appears to codify the self-coupling of the gravitational field. Unfortunately, this object has many undesirable features: it is not a tensor under general coordinate transformations, has no clear physical interpretation, and fails to reveal any equivalence between the coupling of gravity to matter, and gravity to itself. 

We propose an alternative solution to this apparent inconsistency: the action for the graviton is not the Fierz-Pauli action but is instead $S_2$ given by (\ref{newact}), possessing a non-minimally coupled term that vanishes when the (vacuum) background equations are enforced.\footnote{More precisely, $S_2$ is the action for the graviton in a background spacetime with metric in some small neighbourhood of the solutions of the vacuum field equations. We use the term \emph{vacuum} to signify a region without matter; this does not necessarily imply the absence of spacetime curvature.} We shall demonstrate that the energy-momentum tensor of this action is the correct second-order contribution to the equation of motion, and furthermore, that this action provides the starting point for a straightforward energy-momentum self-coupling procedure that generates the Einstein-Hilbert action (modulo surface terms) to \emph{arbitrary order}. We conclude the discussion by extending our formalism to non-vacuum spacetimes.

Throughout the article we employ the abstract index notation \cite{wald}, with lower-case Roman indices indicating a tensor's `slots', and Greek indices serving to enumerate its components in a particular coordinate system. The metric has signature $(-,+,+,+)$, $\kappa\equiv8\pi G/c^4$, and the Riemann and Ricci tensor are defined with the following conventions: $R^a_{\phantom{a}bcd}v^b\equiv2 \nabla_{[c}\nabla_{d]}v^a$, $R_{ab}\equiv R^c_{\phantom{c}acb}$.

\section{The Graviton Action}\label{act}
Contrary to the standard approach, we represent the gravitational field as a perturbation $h^{ab}$ of the \emph{inverse} physical metric $g^{ab}$ from the background $\gb^{ab}$: 
\bea \label{hdef}
g^{ab}= \gb^{ab} + h^{ab}.
\eea
This expression is \emph{exact} in that we have not neglected terms $O(h^2)$; in contrast, the physical metric $g_{ab}=\gb_{ab} - h^{cd}\gb_{ca}\gb_{db}+ O(h^2)$ . Following this convention, we use the contravariant field $h^{ab}$, rather than $h_{ab}$, as the fundamental dynamical variable of the action.\footnote{Any metric theory of gravity will have an ambiguity as to which variable $g \in\{g^{ab}, g_{ab}, \sqrt{-g}g^{ab},\ldots\}$ should be identified as the true ``gravitational field''. Such a distinction is of no physical consequence and is largely unnecessary for a non-perturbative calculation; however for the present discussion we are forced to single out a particular field variable for the expansion $g=\gb + h$. Our aim is to connect gravity to the particle physics notion of a spin-2 field and elucidate a simple energy-momentum self-coupling scheme that generates general relativity; to this end we are required to pick $g\in \{g^{ab}, g_{ab}\}$ as it is only for these that $h$ is a genuine spin-2 field, i.e.\ a symmetric tensor (not a tensor density) with (lowest-order) infinitesimal gauge transformation $\delta h^{ab}= 2\delb^{(a}\epsilon^{b)}$. Fortunately, it is precisely for $g\in \{g^{ab}, g_{ab}\}$ that the necessary energy-momentum self-coupling is its most simple: $h^{ab}t_{ab}$ (see \S\ref{pert}). These considerations provide no criteria for choosing the metric over its inverse as our expansion variable, and while this choice only trivially alters the perturbation theory at first-order ($h^{ab}\leftrightarrow -h_{ab}$) to second-order (the relevant order for $S_2$, $t_{ab}$, and $G^{(2)}_{ab}$) the two definitions of the $h$-field differ by a term of the form $h^{ac}h^{b}_{\phantom{b}c}$. Our choice of $g=g^{ab}$ is preferable for this article because it simplifies the mathematics of the action and energy-momentum tensor. The reason for this is explored in \S\ref{S2deriv}, and stems from the fact that any Lagrangian for pure gravity must contain more factors of $g^{ab}$ than $g_{ab}$ in order that all the derivatives $\partial_a$ be contracted; thus an expansion in $g=g^{ab}$ will be algebraically simpler. Indeed, this observation still holds when coupling gravity to a scalar field $\phi$ or a 1-form $A_a$, and thus taking $g=g^{ab}$ simplifies many of the calculations of the non-vacuum case also (see \S\ref{matter}).} 
In general we will write bars over tensors derived solely from the background geometry, and adopt the usual notational convenience of raising and lowering indices with $\gb^{ab}$ and $\gb_{ab}$.\footnote{The only exception to this rule is the physical metric and its inverse, for which $g^{ab}\ne g_{cd}\gb^{ac}\gb^{db}$, but rather $g^{ab}g_{bc}=\delta^a_c$.}

We posit that the dynamics, energy and momentum of the gravitational field $h^{ab}$, propagating in a background spacetime with metric $\gb_{ab}$, are all determined (to lowest-order) by the following action:
\bea\label{newact}
S_2[\gb^{ab}, h^{ab}]\equiv\frac{1}{2\kappa}\int\!\ud^4x \sqrt{-\gb} h^{ab}(\hat{G}_{abcd}+ \bar{H}_{abcd})h^{cd},
\eea
where 
\bea\nonumber
\hat{G}_{abcd} &\equiv& \half (\gb_{a(c}\gb_{d)b} - \gb_{ab}\gb_{cd})\delb^2 -  \delb_{(c}\gb_{d)(a}\delb_{b)} \\\label{Gdef}
&& {}+ \half\gb_{ab}\delb_{(c}\delb_{d)} + \half\gb_{cd}\delb_{(a}\delb_{b)}
\eea
is a differential operator representing the linearised Einstein tensor (see appendix \ref{calc}) and 
\bea\label{Hdef}
\bar{H}_{abcd}&\equiv &\half\Rb(\gb_{ac}\gb_{db} + \half \gb_{ab}\gb_{cd})  - \Rb_{ab}\gb_{cd}.
\eea
While $\bar{H}_{abcd}$ has no obvious geometric interpretation, we intend to show that its contribution to the action is necessary for the consistency of energy-momentum self-coupling with general relativity. Further motivation for this ansatz is given in section \ref{pert}. 

Naturally, if we are to obtain general relativity without at first assuming it, we must begin by considering the graviton in a \emph{flat} background spacetime. Nevertheless, we will see from the formalism of section \ref{pert} that (provided we use $S_2$ to describe the graviton) energy-momentum self-coupling generates the Einstein-Hilbert action even when the background is not flat; $\gb^{ab}$ need only satisfy the weaker condition
\bea\label{backeq}
\bar{G}_{ab} \equiv \Rb_{ab} - \half \gb_{ab}\Rb  =0.
\eea
While this equation expresses the generality of the analysis that is to follow, it should be stressed that no knowledge of (\ref{backeq}) will be required to assemble the Einstein-Hilbert action order by order: a flat background will serve as a perfectly satisfactory starting point.\footnote{Of course, once the self-coupling procedure is complete, and the Einstein-Hilbert action has been assembled starting from the graviton on a flat background, we will be in a excellent position to justify (\ref{backeq}), as this is precisely the field equation (applied to the background) that we will have derived. With hindsight, then, we can see there was nothing special about our flat-space starting point: we may begin with any \emph{one} solution to (\ref{backeq}) and use energy-momentum self-coupling to derive the action (and field equation) that defines \emph{all} the others.} No matter which background we use, however, it is absolutely crucial that we refrain from inserting this particular metric (or even equation (\ref{backeq})) into the action, thereby reducing $S_2$ to $\frac{1}{2\kappa}\int\!\ud^4x \sqrt{-\gb} h^{ab}\hat{G}_{abcd}h^{cd}$. This is because we will need to be able to perform arbitrary variations of $\gb^{ab}$, not just those consistent with $\bar{R}_{abcd}=0$ or $\bar{R}_{ab}=0$, to construct the energy-momentum tensor for $h^{ab}$. That said, it will be instructive to temporarily ignore this advice so that we may relate $S_2$ to the Fierz-Pauli action.

\subsection{The Fierz-Pauli action}\label{FPaction}
For a flat background, $\bar{H}_{abcd}$ vanishes, and we can choose coordinates $\{x^\alpha\}$ such that $\gb^{\alpha\beta} = \eta^{\alpha\beta}$ and evaluate $S_2$ as a functional of the components $h^{\alpha\beta}$. Integrating by parts and discarding surface terms, we find that $S_2$ reduces to $\frac{-1}{2\kappa}\int\!\ud^4x \mathcal{L}_{\text{FP}}$, where
\bea\nonumber\label{FP}
\mathcal{L}_{\text{FP}} & = & \half \partial_\lambda h_{\alpha\beta}\partial^\lambda h^{\alpha\beta} -  \half\partial_\lambda h \partial^\lambda h -  \partial_\lambda h^{\alpha\beta}\partial_\alpha h^{\phantom{\beta}\lambda}_{\beta}  \\ && {} + \partial_\alpha h \partial_\beta h^{\alpha\beta}
\eea
is the Fierz-Pauli Lagrangian \cite{Pad}.\footnote{Here and elsewhere we use the customary shorthand $h\equiv h^{a}_{\phantom{a}a}\equiv  h^{ab}\gb_{ab}$.} Modulo surface terms and an overall rescaling, $\mathcal{L}_{\text{FP}}$ is the unique specially relativistic Lagrangian for a symmetric tensor field $h^{\alpha\beta}$ that is invariant under the infinitesimal gauge transformation $\delta h^{\alpha\beta}= 2\partial^{(\alpha}\epsilon^{\beta)}$ (see \cite{Pad} for proof); hence it is the Lagrangian for the graviton (massless spin-2 field) in flat spacetime. 

Starting from (\ref{FP}), we can ``covariantize'' $\mathcal{L}_{\text{FP}}$ by making the replacements $\eta_{\alpha\beta} \to \gb_{\alpha\beta} $, $\partial_\alpha \to \delb_\alpha$ and multiplying by $\sqrt{-\gb}$. This process obviously generates a unique manifestly covariant Lagrangian density if $\gb^{ab}$ is flat, as in this case the procedure is equivalent to representing the same Lagrangian in arbitrary coordinates. However, for the purposes of calculating the energy-momentum tensor (via arbitrary variations of $\gb^{ab}$) it will be necessary to generalize $\mathcal{L}_{\text{FP}}$ to arbitrary backgrounds, and for a curved metric the covariantization procedure is ambiguous. To see this, observe that we can transmute the third term of (\ref{FP}) by twice integrating by parts:
\bea\label{trans}
\partial_\lambda h^{\alpha\beta}\partial_\alpha h^{\phantom{\beta}\lambda}_{\beta} &\leftrightarrow& \partial_\alpha h^{\alpha\beta}\partial_\lambda h^{\phantom{\beta}\lambda}_{\beta}. 
\eea
However this equivalence relies on the commutativity of partial derivatives, and does not occur for the covariant derivatives of a curved background; instead, integration by parts yields
\bea\nonumber
\delb_c h^{ab}\delb_a h^{\phantom{b}c}_{b} &\leftrightarrow& \delb_a h^{ab}\delb_c h^{\phantom{b}c}_{b}  - h^{ca}h^{b}_{\phantom{b}c}\Rb_{ab} \\ \label{covamb} && {} - h^{ab}h^{cd}\Rb_{acdb}. 
\eea
Thus we are forced to make a seemingly arbitrary choice: do we to covariantize (\ref{FP}) as written, or should we do so after performing (\ref{trans})? These two possibilities determine Lagrangians which differ by  $h^{ca}h^{b}_{\phantom{b}c}\Rb_{ab}+ h^{ab}h^{cd}\Rb_{acdb}$; they lead to different (first-order) equations of motion if the background is curved,\footnote{The first-order field equation only describes the spacetime perturbations of general relativity if the ambiguous term is covariantized to become $\delb_c h^{ab}\delb_a h^{\phantom{b}c}_{b}$; see \S\ref{fieldeq} and Appendix \ref{calc}.} and determine different energy-momentum tensors even if the background is flat.\footnote{Note that all other terms of $\mathcal{L}_{\text{FP}}$ are invariant under the operation that generated (\ref{trans}) so do not introduce further ambiguity.} This last problem is discussed by Padmanabhan \cite{Pad}, and is one of his many non-trivial objections to the conventional wisdom that general relativity is the unique energy-momentum self-coupled limit of the flat-space massless spin-2 field.

A greater problem than this ambiguity, however, is that neither choice (nor an admixture) leads to general relativity after coupling it to its own energy-momentum. As we shall see in section \ref{pert}, the contribution from $h^{ab}\bar{H}_{abcd}h^{cd}$ is necessary to achieve this, and it is impossible to use the covariantizing ambiguity to produce this tensor because it does not contain $h^{ab}h^{cd}\Rb_{acdb}$. Instead, the presence of $\bar{H}_{abcd}$ represents a rather different coupling ambiguity faced when moving from a flat background to a curved one. Typically we would invoke the Einstein equivalence principal to banish from the action terms coupling matter fields and Ricci tensors; we would argue that, working in locally inertial coordinates about a point $p$, the Lagrangian at $p$ should have the same form as the Lagrangian in flat spacetime. This amounts to a minimal coupling procedure: once we have covariantized a specially relativistic Lagrangian, the job of coupling the field to the gravity is complete. However, while this rule may make sense to curve the background spacetime of a spin-2 field that is ``just another matter-field'' and has nothing to do with gravitation, it is far from clear that the principal should hold for the graviton, for which it was only ever a convenient fiction to think of as a tensor field propagating over a background geometry.

In summary, the Fierz-Pauli action is insufficient to determine $S_2$ for an arbitrary background geometry; the principal of equivalence fails to give a unique solution, and cannot justify all the contributions necessary for an energy-momentum self-coupling procedure consistent with general relativity. However, it was never our aim to construct general relativity from $\mathcal{L}_{\text{FP}}$, and we do not pretend to be able to derive a curved spacetime theory of gravity from purely specially relativistic concepts. $S_2$ will serve as our starting point, and the only significance we shall ascribe $\mathcal{L}_{\text{FP}}$ is that of a special case. 

\subsection{Field equations}\label{fieldeq}
Leaving the Fierz-Pauli action behind, we retrain our attention on  $S_2$ and begin the process of deriving its advertised connection to general relativity. First, we shall calculate the associated field equations. As usual, the equations of motion are derived from the condition that their solutions be stationary configurations of $S_2$ with respect to variations in the dynamical field $h^{ab}$. As we will have no cause to vary $\gb^{ab}$ in the derivation, we can enforce the background equations (\ref{backeq}) immediately and discard $\bar{H}_{abcd}$. Next, observe that $\hat{G}_{abcd}$ is ``self-conjugate'': for any tensor fields $A^{ab}$ and $B^{ab}$ 
\bea\label{selfconj}
\int\!\ud^4x \sqrt{-\gb} A^{ab}\hat{G}_{abcd}B^{cd} = \int\!\ud^4x\sqrt{-\gb} B^{ab}\hat{G}_{abcd}A^{cd},
\eea
provided either $A^{ab}$ or $B^{ab}$ has compact support. Therefore, holding $\gb^{ab}$ constant and performing a variation $\delta h^{ab}$ (a symmetric tensor field with compact support) gives rise to a variation in the action
\bea
\delta S_2 & =&  \frac{1}{\kappa}\int\!\ud^4x \sqrt{-\gb}  \delta h^{ab}\hat{G}_{abcd}h^{cd}.
\eea
As $\hat{G}_{abcd}$ is already symmetric in its first two indices, we can conclude that the equation of motion is
\bea \label{eqmotion1}
\frac{1}{\sqrt{-\gb}}\fd{S_2}{h^{ab}}=\kappa^{-1}\hat{G}_{abcd}h^{cd}=0.
\eea
The centrally important feature of this equation is that $\hat{G}_{abcd}h^{cd} = G^{(1)}_{ab}$, the linear approximation to the Einstein tensor under the inverse metric expansion (\ref{hdef}). This is particularly easy to verify for the special case of a flat background in Lorentzian coordinates, but is shown to hold more generally for vacuum backgrounds in Appendix \ref{calc}. Thus $S_2$ prescribes the correct first-order equation of motion for the graviton. In the next section we show that by adding the energy-momentum tensor $t_{ab}$ of $h^{ab}$ (determined by $S_2$) to the right hand side of (\ref{eqmotion1}) we successfully generate the Einstein field equations correct to \emph{second-order}.\footnote{Of course, the resulting field equation will no longer be a stationary configuration of the action $S_2$. In order that this self-coupled equation of motion can be derived from the principle of stationary action it will be necessary to introduce a third-order correction to the action $S_3$. Naturally, $S_3$ will alter the energy-momentum tensor of $h^{ab}$ by a term $O(h^3)$; however, seemingly by miracle, this will be precisely the \emph{third-order} part of the Einstein field equations. This process continues indefinitely and is explained systematically in \S\ref{pert}. For the moment we content ourselves with exploring the theory to second-order only.}

\subsection{Energy-momentum tensor}\label{EMT}
We will now calculate the energy-momentum tensor of the graviton and relate it to the second-order contribution to the Einstein field equations. We follow Hilbert's prescription and define the energy-momentum tensor as a functional derivative of the action with respect to the (background) metric:
\bea\label{hildef}
t_{ab} \equiv \frac{-1}{\sqrt{-\gb}}\fd{S_2}{\gb^{ab}},
\eea
where $h^{ab}$ (rather than $h_{ab}$ or $h^{a}_{\phantom{a}b}$) is to be held constant when taking this derivative, as this is the field we have taken to be the fundamental dynamical variable.\footnote{In later sections, the tensor written here as $t_{ab}$ will be notated $t^2_{ab}$ to indicate that it is the energy-momentum contribution from the second-order action $S_2$ only. Here we need not make this distinction.} 

As an aside, it is worth contrasting the variational definition (\ref{hildef}) with Noether's (canonical) energy-momentum tensor:
\bea\label{Noetherdef}
t_\mathrm{can}^{\mu\nu} \equiv \frac{\partial \mathcal{L}}{\partial(\partial_{\mu}h^{\alpha\beta})}\partial^\nu h^{\alpha\beta} - \eta^{\mu\nu}\mathcal{L},
\eea
comprising the four conserved currents associated with the invariance of the Lagrangian $\mathcal{L}$ under rigid spacetime translations. The canonical tensor cannot be used in the present discussion for a number of reasons. Firstly, it is not uniquely determined by the action for $h^{ab}$: as it depends directly on the Lagrangian, we are free to alter $t_\mathrm{can}^{\mu\nu}$ by adding a four-divergence to $\mathcal{L}$, without changing either the dynamics of $h^{ab}$ or $S_2$. Secondly, we require a \emph{symmetric} tensor to act as the source for the first-order field equation (\ref{eqmotion1}), but the canonical tensor need not have this property.\footnote{It is true that the canonical tensor can be \emph{made} symmetric by adding to it an identically conserved ``correction'' $\partial_\alpha\phi^{\mu[\nu\alpha]}$, a function of $h^{ab}$ that cancels the antisymmetric part of $t_\mathrm{can}^{\mu\nu}$. However, if we allow this sort of ad hoc adjustment of the energy-momentum tensor, we only exacerbate the problem of non-uniqueness.} Lastly, Noether's definition does not naturally generalize to curved spacetime in such a way that $t_\mathrm{can}^{\mu\nu}$ inherits a \emph{covariant} conservation law \cite{kuch}. None of these issues arise with $t_{ab}$, and in any case our aim has been to connect the coupling between matter and gravity found in general relativity with a perturbative coupling of gravity to itself; it is the Hilbert energy-momentum tensor of matter, not the canonical tensor, that appears in the full Einstein field equations as the gravitational source. For these reasons we discard the canonical tensor and henceforth refer to $t_{ab}$, following Hilbert's prescription (\ref{hildef}), as the energy-momentum tensor of $h^{ab}$.

To begin the calculation of $t_{ab}$, we divide the action into two pieces $S_2 = S_{2G} + S_{2H}$:
\bea
S_{2G}\equiv\frac{1}{2\kappa}\int\!\ud^4x \sqrt{-\gb} h^{ab}\hat{G}_{abcd}h^{cd},\\
S_{2H}\equiv\frac{1}{2\kappa}\int\!\ud^4x \sqrt{-\gb} h^{ab}\bar{H}_{abcd}h^{cd}.
\eea
It will be convenient to perform the functional derivative (\ref{hildef}) on these two components separately. Focusing first on $S_{2G}$, we integrate by parts\footnote{More precisely, one adds to the integrand a divergence of the form $\partial_a(\sqrt{-\gb}[h\delb h]^{a})= \sqrt{-\gb}\delb_a[h\delb h]^{a}$ that alters $S_2$ only by a function of the fields on the boundary (or at infinity) and thus may be neglected for the purposes of functional variation.} so as to remove the second derivatives from the integrand:
\bea
S_{2G}=\frac{-1}{2\kappa}\int\!\ud^4x \sqrt{-\gb} \delb_c h^{ab} \delb_d h^{ef}K_{ab\phantom{c}ef}^{\phantom{ab}c\phantom{ef}d},
\eea
for which we have introduced the abbreviation
\bea\nonumber
K_{ab\phantom{c}ef}^{\phantom{ab}c\phantom{ef}d} &\equiv& \frac{1}{2}\Big(\gb^{cd} \gb_{a(e} \gb_{f)b} -\gb^{cd} \gb_{ab} \gb_{ef}   - 2\delta^{c}_{(e}\gb_{f)(a}\delta^{d}_{b)}\\
&&\phantom{2\Big(}  + \delta^{c}_{(e}\delta^{d}_{f)}\gb_{ab} +\delta^{d}_{(a}\delta^{c}_{b)}\gb_{ef} \Big)\\\nonumber
&=& K_{ba\phantom{c}ef}^{\phantom{ba}c\phantom{ef}d} =K_{ab\phantom{c}fe}^{\phantom{ab}c\phantom{fe}d} = K_{ef\phantom{d}ab}^{\phantom{ef}d\phantom{ab}c}.
\eea
An infinitesimal variation in the inverse background metric $\delta \gb^{ab}$, vanishing on the boundary of the integral, induces a variation in the action 
\bea\nonumber
\delta S_{2G}  &=& \frac{-1}{2\kappa}\int\!\ud^4x \sqrt{-\gb} \Bigg[\delta\gb^{pq} \delb_c h^{ab} \delb_d h^{ef}\Bigg(\frac{\partial K_{ab\phantom{c}ef}^{\phantom{ab}c\phantom{ef}d}}{\partial \gb^{pq}}\\\nonumber
&&{} -\frac{1}{2} \gb_{pq} K_{ab\phantom{c}ef}^{\phantom{ab}c\phantom{ef}d} \Bigg) +4\delb_c h^{ab} C^{(e}_{\phantom{(e}sd}h^{f)s} K_{ab\phantom{c}ef}^{\phantom{ab}c\phantom{ef}d}\Bigg],
\eea
where
\bea\nonumber
C^{a}_{\phantom{a}bc} &\equiv &\tfrac{1}{2}\gb^{ad}\left(\delb_b\delta\gb_{cd}+\delb_c\delta\gb_{bd}-\delb_d\delta\gb_{bc}\right)\\\label{con}
&=&-\tfrac{1}{2}\left(2\delta^a_p \delta^r_{(b}\gb_{c)q}-\gb^{ar}\gb_{bp}\gb_{qc}\right)\delb_r\delta\gb^{pq}
\eea
is the connection that arises from the variation of the covariant derivative: $\nabla_{\gb+\delta\gb}=\delb+C$. We can move the covariant derivatives off $\delta \gb^{pq}$ in the connection term using integration by parts, and arrive at an equation of the form $\delta S_{2G} = \int\!\ud^4x \ \delta\gb^{pq} [\ldots]_{pq}$; the tensor density in square brackets is then the functional derivative we seek:
\bea\nonumber\label{G}
\frac{\kappa}{\sqrt{-\gb}}\fd{S_{2G}}{\gb^{pq}} &=& \frac{-1}{2} \delb_c h^{ab} \delb_d h^{ef}\\\nonumber
&&{} \times\left(\frac{\partial K_{ab\phantom{c}ef}^{\phantom{ab}c\phantom{ef}d}}{\partial \gb^{pq}}- \frac{1}{2} \gb_{pq} K_{ab\phantom{c}ef}^{\phantom{ab}c\phantom{ef}d}\right)\\\nonumber
&&{} - \delb_r\Big( \delb_c h^{ab}\Big(
K_{ab\phantom{c}(p|f|q)}^{\phantom{ab}c}h^{rf}
\\
&&{}+K_{ab\phantom{c}f(p}^{\phantom{ab}c\phantom{f(p}r}h_{q)}^{\phantom{q)}f}
-K_{ab\phantom{cr}f(p}^{\phantom{ab}cr}h_{q)}^{\phantom{q)}f}\Big)\Big).
\eea
Meanwhile, $S_{2H}$ varies by
\bea\nonumber
\delta S_{2H} &=& \frac{1}{2\kappa}\int\!\ud^4x \sqrt{-\gb} \delta\bar{R}_{ab} \\\label{varSH}
&&\phantom{2\kappa x}\times\left(\tfrac{1}{2}\gb^{ab}\left(\tfrac{1}{2}h^2 + h_{cd}h^{cd} \right)- h^{ab} h \right),
\eea
where we have used the background equation (\ref{backeq}) (\emph{after} the variation) to remove the terms proportional to $\bar{R}_{ab}$; these would only be significant if we intended to perform further variations in the metric. Now, because
\bea\nonumber
\delta \bar{R}_{ab}&=& 2\delb_{[c}C^c_{\phantom{c}b]a}\\\nonumber
&=&\left(\half\gb^{rs}\gb_{ap}\gb_{qb} + \half \delta^r_{(a} \delta^s_{b)}\gb_{pq}-\delta^r_p\delta^s_b \gb _{aq}\right)\delb_r\delb_s\delta\gb^{pq},
\eea
when we (twice) integrate by parts to alleviate $\delta \gb^{ab}$ of its covariant derivatives, we generate a second-order differential operator
\bea\label{Rhatdef}
\hat{R}_{pqab}\equiv \tfrac{1}{2}\gb_{a(p}\gb_{q)b}\delb^2  + \tfrac{1}{2}\gb_{pq}\delb_{(a}\delb_{b)} -\delb_{(a} \gb_{b)(p} \delb_{q)},
\eea
with the property
\bea\label{Rhatprop}
\int\!\ud^4x \sqrt{-\gb} \delta\bar{R}_{ab}A^{ab} = \int\!\ud^4x \sqrt{-\gb} \delta\gb^{pq}\hat{R}_{pqab}A^{ab}
\eea
for all $A^{ab}$. Therefore, we can conclude from (\ref{varSH}) that
\bea\nonumber
\frac{\kappa}{\sqrt{-\gb}}\fd{S_{2H}}{\gb^{pq}} &=& \frac{1}{2} \hat{R}_{pqab}\left(\tfrac{1}{2}\gb^{ab}\left(\tfrac{1}{2}h^2 + h_{cd}h^{cd} \right)- h^{ab} h \right).\\\label{H}
\eea

%
Finally, we have only to combine equations (\ref{G}) and (\ref{H}), expand out all the products and derivatives, and assemble the outcome into a formula for $t_{ab}$ as a function of $\delb_{c}h^{ab}$. This is a straightforward but arduous calculation, and as such we chose to complete it with a computer algebra package. The result is
\begin{widetext}
\bea\nonumber
\kappa t_{pq}&=& \tfrac{1}{4}\gb_{pq}\Big(h\delb_a\delb_b h^{ab}+2 h^{ab}\delb_a\delb_b h- 2 h_{ab} \delb^2 h^{ab}-h\delb^2h - \tfrac{1}{2}\delb_a h \delb^a h - \tfrac{5}{2} \delb_{c}h_{ab}\delb^{c}h^{ab} 
+ \delb_{c}h_a^{\phantom{a}b}\delb_b h^{ac} 
\\\nonumber
&&{}+2 \delb_a h \delb_b h^{ab}\Big) + \tfrac{1}{4}h\delb_{(p}\delb_{q)}h -\tfrac{1}{2}h_{pq}\delb^2 h + \tfrac{1}{4}h \delb^2 h_{pq} + h_{a(p}\delb^2h_{q)}^{\phantom{q)}a} - \tfrac{1}{2} h^{ab}\delb_a\delb_b h_{pq}+ \tfrac{1}{2} h_{pq}\delb_a\delb_b h^{ab}\\\nonumber
&&{} - h_{a(p}\delb^b\delb_{q)} h^a_{\phantom{a}b}+ \tfrac{1}{2} h_{ab}\delb_{(p}\delb_{q)} h^{ab} - \tfrac{1}{2} h\delb_a\delb_{(p} h_{q)}^{\phantom{q)}a} + \tfrac{1}{4}\delb_a h \delb^{a}h_{pq} +\tfrac{1}{2}\delb_bh_{ap}\delb^{b}h^a_{\phantom{a}q} - \tfrac{1}{2}\delb_ah_{pq}\delb_bh^{ab} \\\label{t}
&&{} + \tfrac{3}{4}\delb_ph_{ab}\delb_q h^{ab}  - \delb_bh^a_{\phantom{a}(p}\delb_{q)}h_a^{\phantom{a}b}- \tfrac{1}{2}\delb_b h\delb_{(p} h_{q)}^{\phantom{q)}b} +  \tfrac{1}{2} \delb_b h^a_{\phantom{a}p}\delb_a h^b_{\phantom{b}q}.
\eea
It is possible to render this formula rather more manageable by working in a gauge with $\delb_a h^{ab}=0$, $h=0$:
\bea\nonumber
\kappa t_{pq} &=& \gb_{pq}\Big(\tfrac{1}{4}\delb_{c}h_a^{\phantom{a}b}\delb_{b}h^{ac} -\tfrac{5}{8} \delb_{c}h_{ab}\delb^{c}h^{ab} - \tfrac{1}{2}h_{ab} \delb^2 h^{ab}
\Big) + h_{a(p}\delb^2h_{q)}^{\phantom{q)}a} - \tfrac{1}{2} h^{ab}\delb_a\delb_b h_{pq} - h^{bc}\bar{R}_{abc(p} h_{q)}^{\phantom{q)}a}\\\label{tgauge}
&&{} + \tfrac{1}{2} h_{ab}\delb_{(p}\delb_{q)} h^{ab} +\tfrac{1}{2}\delb_bh_{ap}\delb^{b}h^a_{\phantom{a}q} + \tfrac{3}{4}\delb_ph_{ab}\delb_q h^{ab}  - \delb_bh^a_{\phantom{a}(p}\delb_{q)}h_a^{\phantom{a}b} + \tfrac{1}{2} \delb_b h^a_{\phantom{a}p}\delb_a h^b_{\phantom{b}q},
\eea
\end{widetext}
but we will not need this partially gauge-fixed result for this present article.\footnote{Gauge transformations are covered in \S\ref{GT}; we note here only that because $t_{ab}$ is not invariant under the infinitesimal gauge transformation $\delta h^{ab}= 2\delb^{(a}\epsilon^{b)}$, only the first formula (\ref{t}) can be used in all gauges. Although gauge invariance would be a highly desirable property if we intended to argue that $t_{ab}$ was a physically meaningful tensor in full general relativity, it is an impossible request to make of the tensor we seek, which should be proportional to the gauge dependent tensor $G^{(2)}_{ab}$.}

Our task now is to compare $t_{ab}$ with $G^{(2)}_{ab}$ and demonstrate that the energy-momentum self-coupling of $h^{ab}$ (determined by $S_2$) is consistent with general relativity. Details of the calculation of $G^{(2)}_{ab}$ can be found in Appendix \ref{calc}; the conclusion is
\bea\label{works}
G^{(2)}_{ab}=-\kappa t_{ab} + O(h^3),
\eea
and thus, to second-order, the vacuum Einstein field equations are
\bea\label{EOM2}
\hat{G}_{abcd}h^{cd}=\kappa t_{ab}
\eea
as advertised.

As a corollary of (\ref{EOM2}), we can confirm Padmanabhan's observation that general relativity cannot be derived from energy-momentum self-coupling the Fierz-Pauli Lagrangian. Only once the contribution from $\bar{H}_{abcd}$ is included will Einstein's gravity result from an energy-momentum self-coupled graviton. This realisation casts doubt on Mannheim's recent treatment of gravitational energy-momentum \cite{man}, in which a tensor is constructed by applying (\ref{hildef}) to a covariantized Fierz-Pauli Lagrangian, rather than $S_2$.

\section{Perturbative Gravity}\label{pert}
Here we develop the formalism to uncover the root cause of the second-order energy-momentum self-coupling (\ref{EOM2}), and reveal how the  process continues to arbitrary order. The vast majority of this section applies to any metric theory of pure gravity\footnote{We require only that the dynamics are determined by an action that is a coordinate-independent integral of the metric and its derivatives.} and can be generalized to include interactions with matter (see \S\ref{matter}). Only in section \ref{S2deriv} will we commit to general relativity, fix our action $S=S_{\mathrm{EH}}$, the Einstein Hilbert action, and derive the formula (\ref{newact}) for $S_2$. 

We shall concern ourselves with an expansion of the inverse metric $g^{ab}$ about a non-dynamical background $\gb^{ab}$, which is itself an exact solution of the vacuum field equations:
\bea\label{exp}
g^{ab} &= &\gb^{ab} + \lambda h^{ab},
\\\label{BE}
0&=&\fd{S[\gb]}{\gb^{ab}},
\eea
where $\lambda$, a dimensionless expansion parameter, is constant over spacetime.

Following (\ref{exp}), the action of the exact theory $S[g]$ becomes a $\lambda$-dependent functional of $\gb^{ab}$ and $h^{ab}$, which  can be Taylor expanded thusly:
\bea\label{taylor}
S[g]=S[\gb + \lambda h] =  \sum^\infty_{n=0}\lambda^n S_n[\gb,h],
\eea
where $S_n$ is the ``$n^\mathrm{th}$ partial action'' given by 
\bea\label{Sndef}
S_n [\gb,h]=\frac{1}{n!}\left(\partial_\lambda^n S[\gb + \lambda h] \right)_{\lambda=0}.
\eea
The derivative $\partial_\lambda$ acts on each instance of $\lambda h^{ab}$ in the integrand of $S[\gb + \lambda h]$ by Leibniz's law, removing the factor of $\lambda$. The `bare' $h^{ab}$ left behind may still be covered by spacetime derivatives $\partial_a$, but these can be moved onto the remainder of the integrand by integration by parts. This operation generates the usual functional derivative:
\bea \label{parts1}
\partial_\lambda S[\gb +\lambda h] & =& \int\!\ud^4 x h^{ab}(x)\fd{}{\gb^{ab}(x)} S[\gb + \lambda h].
\eea
In truth, the left hand side of this equation differs from the right by the surface term $\int\!\ud^4x \partial_a J^a$ created when integrating by parts. As this is only a functional of the fields on the boundary (or as $x^{\mu}\to \infty$ if the integral of $S$ runs over the entire manifold) it will not contribute to  equations of motion or energy-momentum tensors, the calculation of which are dependent only on variations of the field that vanish on the boundary (or have compact support). Hence these surface terms may be neglected for our present purposes.

It follows from the repeated application of (\ref{parts1}) that
\bea \partial_\lambda^n S[\gb +\lambda h] & =& \left[\int\!\ud^4 x h^{ab}\fd{}{\gb^{ab}}\right]^n S[\gb + \lambda h],
\eea
and thus the partial actions (\ref{Sndef}) are given by
\bea\label{parts2} S_n [\gb,h]&=& \frac{1}{n!} \left[\int\!\ud^4 x h^{ab}\fd{}{\gb^{ab}}\right]^n S[\gb].
\eea
An important consequence of this relation is that, using $S_2$ as our starting point, we can generate the entire set of partial actions $\{S_n: n \ge 3\}$ by calculating
\bea
S_n [\gb,h]&=& \frac{2}{n!} \left[\int\!\ud^4 x h^{ab}\fd{}{\gb^{ab}}\right]^{n-2} S_2[\gb, h],
\eea
which is possible provided $S_2$ is known in a \emph{neighbourhood} of whichever particular background (a solution of (\ref{BE})) we are interested in. Note that the first two partial actions do not contribute to the dynamics of $h^{ab}$: $S_0=S[\gb]$ is manifestly independent of $h^{ab}$, and $S_1$ vanishes once the background equation (\ref{BE}) has been enforced. We conclude, therefore, that $S_2$ contains all the information necessary to reconstruct the ``dynamical'' part of the action
\bea\label{dynact}
S_\mathrm{dyn}[\gb,h] \equiv \sum^\infty_{n=2}\lambda^n S_n[\gb,h],
\eea
which itself contains all the dynamical information of the full action $S$. This is absolutely key to the calculations of section \ref{act}, in which we saw the first consequence of this reconstruction process, the recovery of the second-order equation of motion from an action that one would expect to encode only first-order dynamics.

\subsection{Field Equations}
In general, we could let $\lambda$ be a free parameter and, on demanding $\delta S[g]/\delta g^{ab}=0$ for fixed $\gb^{ab}$, derive a $\lambda$-dependent equation of motion $E_\lambda[\gb, h]=0$ for our dynamical field $h^{ab}$. Any $h^{ab}$ that solved this equation would correspond to a metric $g^{ab}=\gb^{ab} +\lambda h^{ab}$ that solved the field equations \emph{exactly}.\footnote{It is advisable to set $\lambda=1$ before attempting to solve $E_\lambda[\gb, h]=0$, as this constant can always be absorbed into the magnitude of $h^{ab}$. Although this refinement was convenient for \S\ref{act}, here we shall keep $\lambda$ as it provides a simple method for tracking the powers of $h^{ab}$ in expressions and is useful as a variable for differentiation.} However, if we are interested in approximating small variations of the metric (i.e.\ the limit $\lambda h^{ab}\to 0$) we can choose some order $N$ to which we want the equation of motion to hold:
\bea \label{eom1}
\fd{ S[g]}{{g^{ab}}}= O(\lambda^{N+1}).
\eea
This is equivalent to
\bea\label{eom12}
\frac{1}{\lambda}\fd{S^{N+1}_\mathrm{dyn}[\gb,h]}{h^{ab}} = O(\lambda^{N+1}),
\eea
where $S^{N+1}_\mathrm{dyn}$ is defined by discarding from $S_\mathrm{dyn}$ those terms that can be neglected in (\ref{eom1}): 
\bea\label{action2}
S^{N+1}_\mathrm{dyn}[\gb,h] \equiv \sum^{N+1}_{n=2}\lambda^n S_n[\gb,h].
\eea
We shall adopt this ``$N^{\mathrm{th}}$-order approximation'' picture for the development of our formalism, as we can always write $N=\infty$ if we wish to discuss the exact theory. 

For the sake of continuity with the previous section, we introduce the notation
\bea\label{Gdef2}
\left. \fd{S_2[\gb,h]}{h^{ab}}\right|_{\delta S[\gb]/\delta \gb^{ab}=0}\equiv \kappa^{-1} \sqrt{-\gb}\hat{G}_{abcd} h^{cd},
\eea
where, because $S_2$ is second-order in $h^{ab}$, $\hat{G}_{abcd}$ will be a linear differential operator dependent only on $\gb^{ab}$.\footnote{The operator $\hat{G}_{abcd}$ defined here coincides with the definition in (\ref{Gdef}) once $S=S_{\mathrm{\mathrm{EH}}}$ has been fixed. This is shown in \S\ref{S2deriv} by deriving $S_2$.} The equation of motion (\ref{eom12}) now takes the form
\bea\label{eom2}
\lambda \hat{G}_{abcd} h^{cd} = - \frac{\kappa}{\lambda\sqrt{-\gb}}\fd{}{h^{ab}}\sum^{N+1}_{n=3}\lambda^n S_n[\gb,h],
\eea
where it should be taken as given that terms $O(\lambda^{N+1})$ have been neglected. This is the $N^\mathrm{th}$-order approximation to the equation of motion for $h^{ab}$ that is consistent with the dynamics of $g^{ab}$ prescribed by the action $S$. The first-order contribution has been separated from the sum so as to evoke the picture of a wave equation $\lambda \hat{G}_{abcd} h^{cd} =0$ with a source. In the next section we will see that the source term on the right of (\ref{eom2}) is indeed the energy-momentum tensor of the field $h^{ab}$, neglecting terms $O(\lambda^{N+1})$. 

\subsection{Energy-momentum tensor}\label{EMtensor}
First we shall demonstrate that the dynamical part of the action (\ref{dynact}) can be generated from $S_2$ by a simple energy-momentum self-coupling procedure. Observe that, as a consequence of (\ref{parts2}), we have
\bea
S_n[\gb,h] = \frac{1}{n}\int\!\ud^4 x h^{ab}\fd{S_{n-1}[\gb,h]}{\gb^{ab}}. 
\eea
Defining the $n^\mathrm{th}$ partial energy-momentum tensor $t^n_{ab}$ by applying Hilbert's prescription to the $n^\mathrm{th}$ partial action,
\bea\label{emt}
t^n_{ab} \equiv \frac{-1}{\sqrt{-\gb}}\fd{ S_n[\gb,h]}{\gb^{ab}},
\eea
we conclude that
\bea\label{emcoupling}
S_n[\gb,h] = \frac{-1}{n}\int\!\ud^4 x \sqrt{-\gb}h^{ab}t^{n-1}_{ab}.
\eea
This makes manifest the energy-momentum self-coupling procedure that allows us to generate the dynamical part of the action (\ref{dynact}) to arbitrary order, given only $S_2$. The $n^{\mathrm{th}}$ partial action is nothing more than the integral of the contraction of $h^{ab}$ with the energy-momentum tensor of the previous partial action (divided by $-n$). The dynamical part of the action is therefore given by
\bea\nonumber
S^{N+1}_{\mathrm{dyn}}[\gb,h] &=& \lambda^2 S_2[\gb, h]\\\label{tinaction}&&{} - \int \ud^4x \sqrt{-\gb}h^{ab}\sum^{N}_{n=2}\frac{\lambda^{n+1} t^{n}_{ab}}{n+1}.
\eea
Note that, for the particular case of general relativity ($S=S_\mathrm{EH}$), the background equation (\ref{backeq}) also sets $S_0=0$, thus $S_\mathrm{dyn}=S_\mathrm{EH}$ (modulo surface terms) and the energy-momentum self-coupling procedure recovers the \emph{entire} action of the full theory, not just the dynamical part. 

Because of factors of $n+1$ dividing each $t_{ab}^{n}$ in (\ref{tinaction}), it is not the case that in the action $h^{ab}$ couples directly to its ($N^\mathrm{th}$-order) total energy-momentum tensor, given by
\bea\label{Tdef}
T^N_{ab} \equiv \frac{-1}{\sqrt{-\gb}}\fd{S^N_\mathrm{dyn}}{\gb^{ab}}=  \sum^{N}_{n=2}\lambda^n t^{n}_{ab}.
\eea
Instead, the numerical denominators account for the $n+1$ factors of $h^{ab}$ in $h^{ab}t_{ab}^{n}$, and ensure that the equations of motion do indeed have $T^N_{ab}$ as the source. To prove this, note that for any symmetric field $l^{ab}$ (vanishing on the boundary, or with compact support) we have
\bea \nonumber
\int\!\ud^4 x l^{ab}\fd{S_n[\gb,h]}{h^{ab}}&=& \int\!\ud^4 x \frac{l^{ab}}{n!}\fd{}{h^{ab}}\left(\partial_\lambda^n S[\gb + \lambda h]\right)_{\lambda=0} \\ \nonumber 
&=&\tfrac{1}{n!}\left(\partial_\mu\left(\partial_\lambda^n S[\gb + \lambda (h +\mu l)]\right)_{\lambda=0} \right)_{\mu = 0}\\ \nonumber
&=&\tfrac{1}{n!}\left(\partial_\lambda^n \partial_\mu S[\gb + \lambda (h +\mu l)]\right)_{\lambda=\mu=0} \\ \nonumber
&=&\tfrac{1}{n!}\left(\partial_\lambda^n \left( \lambda \partial_\alpha S[\gb + \lambda h +\alpha l]\right)\right)_{\lambda=\alpha=0},
\eea
where $\alpha \equiv \lambda \mu \Rightarrow \partial_\mu = \lambda\partial_\alpha $. Thus, 
\bea \nonumber
\int\!\ud^4 x l^{ab}\fd{S_n[\gb,h]}{h^{ab}}&=& \tfrac{1}{n!}\Big(
\lambda \partial_\lambda^n   \partial_\alpha S[\gb + \lambda h +\alpha l]
\\\nonumber &&\phantom{\tfrac{1}{n!}}+ n \partial_\lambda^{n-1} \partial_\alpha S[\gb + \lambda h +\alpha l] 
\Big)_{\lambda=\alpha=0} \\ \nonumber
&=& \tfrac{1}{(n-1)!}\left(\partial_\alpha\partial_\lambda^{n-1} S[\gb + \lambda h +\alpha l] 
\right)_{\lambda=\alpha=0} \\ \nonumber
&=&\left( \partial_\alpha S_{n-1}[\gb + \alpha l, h]\right)_{\alpha=0} \\ \label{Sgh-1}
&=& \int\!\ud^4 x l^{ab}\fd{S_{n-1}[\gb,h]}{\gb^{ab}}.
\eea
Hence we have the following important result:
\bea\label{Sgh}
\fd{S_n[\gb,h]}{h^{ab}} =\fd{S_{n-1}[\gb, h]}{\gb^{ab}}.
\eea
Or, using definition (\ref{emt}),
\bea
\fd{S_n[\gb,h]}{h^{ab}} = -\sqrt{-\gb}t^{n-1}_{ab}.
\eea
Therefore the equation of motion (\ref{eom2}) takes on the form
\bea
\lambda \hat{G}_{abcd} h^{cd} = \kappa \lambda^{-1}
\sum^{N+1}_{n=3}\lambda^n t^{n-1}_{ab},
\eea
or, recalling (\ref{Tdef}),
\bea\label{eom3}
\lambda \hat{G}_{abcd} h^{cd} = \kappa T^N_{ab}.
\eea
We have derived the relation we sought, demonstrating that any metric theory of pure gravity can be formulated as a first-order wave equation with its own energy-momentum tensor as a source. For every $N \ge 1$, we can derive the equation of motion (\ref{eom3}) by applying the variational principle to the action $S^{N+1}_{\mathrm{dyn}}$; the left hand side is the wave equation for the linearised theory, and the right hand side is the energy-momentum tensor prescribed by the action $S^N_{\mathrm{dyn}}$. This energy-momentum tensor is, to some extent, incomplete: it does not include the $O(\lambda^{N+1})$ contribution from the highest-order partial action $S_{N+1}$. This contribution could be calculated, if so desired, and added by hand to the field equations (\ref{eom3}) so that the right hand side read $\kappa T^{N+1}_{ab}$, but this equation would no longer be a stationary configuration of the action $S^{N+1}_{\mathrm{dyn}}$. To remedy this, we could introduce a correction to the action $\lambda^{N+2}S_{N+2}$ that would generate the extra term in the equation of motion; the appropriate functional is given by (\ref{emcoupling}) and couples $h^{ab}$ to the highest-order partial energy-momentum tensor $t_{ab}^{N+1}$. But now once again the energy-momentum tensor $T^{N+1}_{ab}$ is incomplete, and we can apply this same line of reasoning anew. So long as there is no $N$ for which $t^N_{ab}$ vanishes identically, this process can continue indefinitely, and as $N\to \infty$ the exact field equations are recovered, along with the action $S_\mathrm{dyn}=S -S_0 - \lambda S_1$.

All that remains is to connect our formalism to the specific results of the previous section. For the sake of completeness, however, we shall first discuss the gauge symmetries of the theory, and deduce the conservation law for $T^{N+1}_{ab}$.

\subsection{Gauge transformations}\label{GT}
Because the action $S[g]$ is a coordinate-system independent integral, any diffeomorphism $\phi: \mathcal{M}\to\mathcal{M}$ gives rise to a gauge transformation of the theory through the action of $\phi^*$, the map comprising the pullback of $\phi$ on covector indices and the pushforward of $\phi^{-1}$ on vector indices:
\bea\label{diffinv}
S[\phi^* g]= S[g].
\eea
Taylor expanding both sides about $\gb^{ab}$ and applying the background equation reveals the gauge invariance of the dynamical part of the action:
\bea\label{gaugeS}
S^{N+1}_{\mathrm{dyn}}[\gb,h^\prime] = S^{N+1}_{\mathrm{dyn}}[\gb,h], 
\eea
where
\bea
\lambda h^{\prime ab} \equiv \phi^* g^{ab} - \gb^{ab}.
\eea
In the context of an $N^\mathrm{th}$-order approximation, we must insist that $\phi^* = 1 + O(\lambda)$, otherwise these transformations will map the small metric fluctuations $\lambda h^{ab}$ onto fluctuations comparable in magnitude to $\gb^{ab}$. We can write a general diffeomorphism of this form as $\phi^*  = e^{\lambda \mathcal{L}_\xi}$, where $\mathcal{L}_\xi$ is the Lie derivative along a vector field $\xi^a = O(1)$. The gauge transformations of the theory are hence given by
\bea\nonumber
h^{ab}&\to& h^{\prime ab} =  h^{ab} + \delta h^{ab},\\ \label{deltah}
\delta h^{ab} &\equiv& \lambda ^{-1}\sum_{n=1}^{N}\frac{(\lambda \mathcal{L}_\xi)^n}{n!} \gb^{ab} + \sum_{n=1}^{N -1}\frac{(\lambda \mathcal{L}_\xi)^n}{n!}h^{ab},
\eea
where we have discarded all terms $O(\lambda^{N})$, as these will only contribute terms $O(\lambda^{N+1})$ to the equation of motion, and terms $O(\lambda^{N+2})$ to $S^{N+1}_\mathrm{dyn}$. If we wish we can let $\xi^a=\epsilon^a$, an infinitesimal vector field, and derive the infinitesimal gauge transformation
\bea
\delta h^{ab} =\begin{cases} \mathcal{L}_\epsilon\left(\gb^{ab} +\lambda h^{ab}\right) & N\ge 2,\\
-2 \delb^{(a}\epsilon^{b)} &N=1.
\end{cases}
\eea
Because these gauge transformations (infinitesimal or otherwise) are symmetries of $S^{N+1}_\mathrm{dyn}$, they map solutions of the equation of motion (\ref{eom3}) to other solutions. We can therefore use the equation of motion to deduce the transformation law for $T_{ab}^N$:
\bea
\delta T^N_{ab} \equiv T^{N}_{ab}[\gb, h^\prime] - T^{N}_{ab}[\gb, h] = \frac{\lambda}{\kappa} \hat{G}_{abcd} \delta h^{cd}.
\eea
This verifies the earlier remark that the energy-momentum tensor is  gauge dependent, except in the trivial case $N=1$, for which $T^N_{ab}=0$ by definition. It may come as a surprise that the energy-momentum tensor does not inherit the gauge invariance of the action from which it was derived. It should be stressed, however, that $S^{N+1}_{\mathrm{dyn}}$ is not \emph{identically} gauge invariant: the relation (\ref{gaugeS}) is only true when the background equation is obeyed. For general $\gb^{ab}$, the diffeomorphism invariance of $S[g]$ only furnishes the gauge transformation law $\delta S^{N+1}_{\mathrm{dyn}} = -\lambda \delta S_1$, the right-hand side of which has a non-vanishing energy-momentum tensor responsible for the variation in $T^{N}_{ab}$. Equivalently, the gauge dependence of $T^N_{ab}$ can be seen to result from the non-commutativity of gauge transformations and the functional derivative $\delta/\delta \gb^{ab}$ used to define $T^N_{ab}$ \cite{Mag}; these operations would only commute if the gauge invariance of $S^{N+1}_{\mathrm{dyn}}$ extended to a neighbourhood of the solutions of the background equation, rather than being confined to the solutions themselves.

\subsection{Conservation law}\label{consL}
It should be expected that $S^{N+1}_{\mathrm{dyn}}[\gb,h]$ inherits the diffeomorphism invariance of $S[g]$, and that this symmetry endows the energy-momentum tensor with a covariant conservation law with respect to the background metric. The derivation proceeds in close analogy to the proof of $\nabla^a T_{ab}^{\mathrm{matter}}=0$ from general relativity.

We again appeal to the diffeomorphism invariance of the action (\ref{diffinv}) but this time expand $S[g]$ about $\gb^{ab}$ (a solution of the background equation) and $S[\phi^* g]$ about $\phi^* \gb^{ab}$ (which will also be a solution). The result,
\bea\label{diffinv2}
S^{N+1}_{\mathrm{dyn}}[\phi^* \gb, \phi^* h] = S^{N+1}_{\mathrm{dyn}}[\gb,h],
\eea
affirms that $S^{N+1}_{\mathrm{dyn}}$ is diffeomorphism invariant.\footnote{Note that diffeomorphism invariance is equivalent to being independent of coordinate system, and is a distinct property from gauge invariance as defined in \S \ref{GT}.} Now let $\phi$ be an infinitesimal diffeomorphism: $\phi^* =1 + \mathcal{L}_\epsilon$ for an arbitrary infinitesimal vector field $\epsilon^a$ with compact support. Then (\ref{diffinv2}) becomes
\bea
0 = \int\!\ud^4 x \left[\fd{S^{N+1}_{\mathrm{dyn}}}{\gb^{ab}}\mathcal{L}_\epsilon \gb^{ab} + \fd{S^{N+1}_{\mathrm{dyn}}}{h^{ab}}\mathcal{L}_\epsilon h^{ab} \right].
\eea
Clearly the second term vanishes (to $O(\lambda^{N+1})$) if $h^{ab}$ solves the equation of motion (\ref{eom3}), and thus
\bea \nonumber
0 &=& \int\!\ud^4 x \fd{S^{N+1}_{\mathrm{dyn}}}{\gb^{ab}} \delb^a\epsilon^b  + O(\lambda^{N+2})\\
&=&  \int\!\ud^4 x \sqrt{-\gb} \epsilon^b \delb^a T^{N+1}_{ab} + O(\lambda^{N+2}).
\eea
As this equation holds for any $\epsilon^a$ it follows that
\bea \label{cons}
\delb^a T^{N+1}_{ab} = 0
\eea
is valid up to and including $O(\lambda^{N+1})$. Because this relation holds whenever $h^{ab}$ solves its equation of motion, and because gauge transformations map solutions to solutions, the conservation law is gauge invariant.

It is important to recognize that (\ref{cons}) applies to the $(N+1)^{\mathrm{th}}$-order energy-momentum tensor: this is the highest-order approximation to the energy-momentum tensor that can be constructed from our truncated action $S^{N+1}_{\mathrm{dyn}}$, and is a better approximation than the tensor $T^N_{ab}$ which features in the equations of motion appropriate to this order. Of course, the conservation law for $T^N_{ab}$ follows from (\ref{cons}) by discarding the highest-order term, and ensures the consistency of the equation of motion (\ref{eom3}) with the identity $\delb^a\hat{G}_{abcd}h^{cd}=0$, which holds for all $h^{ab}$ once the background equation has been enforced.

\subsection{Constructing the graviton action}\label{S2deriv}
It is now time to close the circle of our discussion and connect the abstract formalism to our earlier calculation. We shall derive here the graviton action $S_2$, the ansatz of section \ref{act}, by applying the perturbative formalism to the particular case
\bea
S[g]=\frac{1}{\kappa} \int\!\ud^4x \sqrt{-g} R\equiv S_\mathrm{EH}[g],
\eea
the Einstein-Hilbert action. To proceed, we will use equation (\ref{parts2}) to derive $S_1$, and then $S_2$, by successive functional derivatives $\delta/\delta\gb^{ab}$ acting on $S_\mathrm{EH}[\gb]$. The first derivative generates
\bea
S_1[\gb,h]= \frac{1}{\kappa} \int\!\ud^4x \sqrt{-\gb}\bar{G}_{ab} h^{ab},
\eea
which of course vanishes for all $h^{ab}$ when $\gb^{ab}$ solves the background equation $\bar{G}_{ab}=0$. A second variation in $\gb^{ab}$ gives rise to 
\bea\nonumber
\delta S_1 &=& \frac{1}{\kappa} \int \!\ud^4x \sqrt{-\gb} \big[ \delta \bar{R}_{ab}\big(h^{ab} - \half h \gb^{ab}\big)\\\nonumber
&&{} + \delta \gb^{cd}\half\big(h_{cd} \bar{R}- h \bar{R}_{cd}- \gb_{cd} \bar{G}_{ab}h^{ab}  \big) \big].
\eea
Replacing $\delta \bar{R}_{ab} \to \delta \gb^{cd} \hat{R}_{cdab}$ in accordance with (\ref{Rhatprop}), we determine $\delta S_1/\delta \gb^{ab}$ and assemble
\bea\nonumber
S_2&=&\frac{1}{2}\int\!\ud^4 x h^{cd}\fd{S_1}{\gb^{cd}}\\\nonumber
&=& \frac{1}{2\kappa} \int \!\ud^4x \sqrt{-\gb} \big[  h^{cd}\hat{R}_{cdab}\big(h^{ab} - \half h \gb^{ab}\big)\\\nonumber
&&\phantom{2\kappa\ud^4x \sqrt{-\gb} } + \half h^{cd}\big(h_{cd} \bar{R}- h \bar{R}_{cd}- \gb_{cd} \bar{G}_{ab}h^{ab} \big)\big]\\
&=& \frac{1}{2\kappa}\int\!\ud^4x \sqrt{-\gb} h^{ab}(\hat{G}_{abcd}+ \bar{H}_{abcd})h^{cd}.
\eea
In the last line we referred to the definitions (\ref{Gdef}) and (\ref{Hdef}), and made use of the identity
\bea
\hat{R}_{abef}(\delta^e_{c} \delta^f_{d} - \half \gb^{ef}\gb_{cd})\equiv \hat{G}_{abcd}.
\eea
This completes the derivation of the graviton action (\ref{newact}) and confirms that it can be used as the starting point of an energy-momentum self-coupling procedure (\ref{emcoupling}) that generates the Einstein field equations and the Einstein-Hilbert action (modulo surface terms) to arbitrary order.

The preceding calculation helps to reveal the advantage of using $h^{ab}$, a perturbation in the \emph{inverse} metric, as our fundamental degree of freedom. Had we instead taken the usual approach, expanding $g_{ab}= \gb_{ab} + \lambda \mathfrak{h}_{ab}$ and taking $\mathfrak{h}_{ab}$ as fundamental, the perturbative formalism would have unfolded identically but for the placement of indices. However, the calculation of $S_2$ from $S_\mathrm{EH}$ would have differed dramatically. The Lagrangian of $S_1$ would instead be proportional to $\bar{G}^{ab} \mathfrak{h}_{ab}$, and because the Ricci tensor is naturally covariant, the variation of $\bar{G}^{ab}=\bar{R}_{cd}\gb^{ca}\gb^{db}-\half\bar{R}_{cd}\gb^{cd}\gb^{ab}$ under $\delta\gb^{ab}$ would have been complicated by the extra two factors of $\gb^{ab}$ on the first term, compared to the relevant tensor in our approach: $\bar{G}_{ab}= \bar{R}_{ab} - \half \bar{R}_{cd}\gb^{cd}\gb_{ab}$. This trend continues at every order; the $\mathfrak{h}_{ab}$ convention leads to a greater proliferation of terms in each partial energy-momentum tensor because the Lagrangian of $S_n$ has the form $(\delb_a)^2(\mathfrak{h}_{ab})^n$ so must be contracted with a further $n+1$ factors of $\gb^{ab}$ to render it a scalar.\footnote{There are of course the instances of $\gb^{ab}\partial_c \gb_{de}$ in each $\delb_a$, but these occur equally in either convention.} Each of these metric factors generates a term in the partial energy-momentum tensor, and thus act as compound interest for the process of energy-momentum self-coupling. In comparison, our convention leads to Lagrangians of the form $(\delb_a)^2(h^{ab})^n$, which only need only $n-1$ additional factors of $\gb_{ab}$.\footnote{This does not mean that \emph{all} terms in such a Lagrangian will contain only $n-1$ additional factors of $\gb_{ab}$; there will often be cases in which $\gb^{ab}$ is contracted with $(\delb_a)^2$ and thus  $n+1$ factors of the metric (and its inverse) will be present. These cases only represent a small proportion of all possible terms, particularly as $n$ becomes large, and are no worse than the terms afforded by the $\mathfrak{h}_{ab}$ convention.} Clearly the inefficiency of the $\mathfrak{h}_{ab}$ approach stems from the natural covariance of derivative operators ($\partial_a$ or $\delb_a$) and curvature tensors; the advantages of the contravariant expansion $g^{ab}=\gb^{ab}+ h^{ab}$ are therefore not peculiar to the Einstein Hilbert action, and are expected to be even more distinguished in higher derivative theories of gravity.

\section{Matter}\label{matter}
To avoid over-complicating our discussion, we have so far focused exclusively on \emph{pure gravity}. Here we will go some way to remedy this simplification, and generalize the formalism of the previous section to include the perturbations of matter fields, and the effects of non-vacuum backgrounds.

In the most general case, let the action $S$ be a functional of $g^{ab}$ and a generic matter field $\Psi^A$, where $A$ will serve as a placeholder for any number of internal or spacetime indices. We then expand $S$ about a background $(\gb^{ab}, \Psib^A)$ as follows:
\bea
\label{Matterexp1}
g^{ab} &=& \gb^{ab} + \lambda h^{ab},\\\label{Matterexp2}
\Psi^A &=& \Psib^A + \lambda \psi^A,\\
\Rightarrow \quad S[g, \Psi]&=& \sum^\infty_{n=0}\lambda^n S_n[\gb,h,\Psib,\psi],
\eea
where  $\gb^{ab}$ and $\Psib^A$ satisfy the background equations 
\bea
\label{BEM}
\fd{S[\gb, \Psib]}{\gb^{ab}} =0,\quad \fd{S[\gb, \Psib]}{\Psib^A}=0.
\eea
As before, each partial action can be calculated from the partial action at the previous order; with matter included, the appropriate recurrence relation is
\bea\label{mattercouple}
S_n = \frac{-1}{n}\int\!\ud^4x\sqrt{-\gb}\left(h^{ab} t^{n-1}_{ab} + \psi^A j_A^{n-1}\right),
\eea
where
\bea\label{tandj}
t^n_{ab} \equiv \frac{-1}{\sqrt{-\gb}}\fd{ S_n}{\gb^{ab}}, \quad j_A^n\equiv \frac{-1}{\sqrt{-\gb}}\fd{ S_n}{\Psib^A}.
\eea
There are two aspects of this coupling scheme that differ from pure gravity. The first is immediately apparent: the $h^{ab}t_{ab}$ term has been joined by an analogous coupling between matter fluctuations $\psi^A$ and its ``source current'' $j_A$. The second difference is hidden within the definitions of $t_{ab}$ and $j_A$; because the $\{S_n\}$ now represent the partial actions for gravity and matter together, $h^{ab}t_{ab}$ and $\psi^A j_A$ are no longer just self-couplings, and will in general contain terms coupling $h^{ab}$ to $\psi^A$. In particular, $t^n_{ab}$ should now be interpreted as the ($n^\mathrm{th}$-order) energy-momentum tensor due to \emph{all} the fields: $h^{ab}$, $\psi^A$, and the background matter $\Psib^A$.

Proceeding as before, we can now demand that the dynamical fields $h^{ab}$ and $\psi^A$ solve the field equations of the action
$S_\mathrm{dyn}^{N+1} = \sum^{N+1}_{n=2}\lambda^n S_n,$
and generate approximate solutions of the exact field equations (prescribed by $S$) accurate to $O(\lambda^N)$. Instead of using the definition (\ref{Gdef2}) for $\hat{G}_{abcd}$, we write the general form of $S_2$, modulo surface terms, as
\bea\nonumber
S_2&=&\frac{1}{2}\int\!\ud^4 x \sqrt{-\gb}\Big(h^{ab}\hat{G}_{abcd}h^{cd}/\kappa \\
&&\phantom{2\ud^4 x } - 2h^{ab}\hat{I}_{abA}\psi^A+ \psi^A\hat {W}_{AB}\psi^B \Big),
\eea
once the background equations (\ref{BEM}) have been enforced. In the above equation, $\hat{G}_{abcd}$, $\hat{I}_{abA}$, and $\hat{W}_{AB}$ are linear operators that depend only on background fields, $\hat{G}_{abcd}$ and  $\hat{W}_{AB}$ are self-conjugate, in the sense given by (\ref{selfconj}), and $\hat{I}_{abA}$ is conjugate to $\hat{I}_{Aab}^\dagger$:
\bea
\int\!\ud^4x \sqrt{-\gb} A^{ab}\hat{I}_{abA}B^{A} = \int\!\ud^4x\sqrt{-\gb}B^{A}\hat{I}^\dagger_{Aab} A^{ab},
\eea
for all $A^{ab}$ or $B^{ab}$, provided one has compact support. These definitions lead to equations of motion, accurate to $O(\lambda^N)$, as follows:
\bea
\lambda \hat{G}_{abcd} h^{cd}   &= &\kappa T^N_{ab} + \lambda \kappa \hat{I}_{abA}\psi^A,\\
\lambda \hat{W}_{AB} \psi^B  &=& J^N_A + \lambda \hat{I}^{\dagger}_{Aab}h^{ab},
\eea
where
\bea
T^N_{ab} \equiv \sum^{N}_{n=2}\lambda^n t^{n}_{ab}, \quad J^N_A \equiv \sum^{N}_{n=2}\lambda^n j^{n}_A.
\eea

Although this formalism is quite general, it is probably too general to be usefully employed. Indeed, the complications involved in describing matter as a background field \emph{and} a dynamical perturbation generally serve to obscure the physical interpretation of the mathematics. An interesting example of this occurs when one tries to rederive $\delb^a T^{N+1}_{ab} =0$ by applying the argument of section \ref{consL}. The result that now follows is
\bea
\delb^a T^{N+1}_{ab} = \frac{1}{2 \sqrt{-\gb}} \fd{}{\epsilon^b} \int\!\ud^4 x \sqrt{-\gb} J^{N+1}_A \mathcal{L}_\epsilon \Psib^A,
\eea
the physical interpretation of which is far from clear. Rather than continue with this formulation in its full generality, it will therefore be more instructive to examine two special cases. First, we set $\Psib^A=0$ and consider small matter fields $\lambda\psi^A$ interacting with $\lambda h^{ab}$. Second, by setting $\psi^A=0$ we can study the effect of a background matter field $\Psib^A$ on the propagation of the graviton. In principal, one could reach these special cases starting from the formalism we have just described, but it will be simpler and more illuminating to build them up from scratch.

\subsection{Matter perturbations}
In a region where the matter fields are small enough that their effects on spacetime curvature can be described by small perturbations $\lambda h^{ab}$ in the inverse metric, we can model the dynamics by taking $\Psib^A=0$, and describe the matter field using $\lambda \psi^A$ alone.  As it is often the case for gravitational theories, let us suppose that the action $S$ is the sum of a gravitational action $S_\mathrm{g}$ and a matter action $S_\Psi$:
\bea
S[g,\Psi]= S_\mathrm{g}[g]+ S_\Psi[g, \Psi].
\eea
Moreover, for the sake of simplicity, we take $\Psi^A$ to be a \emph{free} field: 
\bea\label{free}
S_\Psi[g,\lambda \Psi]= \lambda^2 S_\Psi[g,\Psi]\quad \forall \ g^{ab}, \Psi^A.
\eea
This assumption will mean that the perturbative expansion of $S$ can be described by an energy-momentum coupling procedure only. To see this explicitly, we expand the action about a background $(\gb^{ab},0)$:
\bea
S[\gb + \lambda h, \lambda \psi]= \sum^\infty_{n=0}\lambda^n\left(S_{\mathrm{g}n}[\gb,h]+S_{\Psi n}[\gb,h,\psi]\right),
\eea
where each gravitational partial action
\bea\nonumber
S_{\mathrm{g}n}[\gb,h]&=& \frac{1}{n!}\left(\partial_\lambda^n S_\mathrm{g}[\gb + \lambda h] \right)_{\lambda=0}\\
&=&\frac{1}{n!}\left[\int\!\ud^4 x h^{ab}\fd{}{\gb^{ab}}\right]^n S_\mathrm{g}[\gb],
\eea
much as before, and the matter partial actions
\bea\nonumber
S_{\Psi n}[\gb,h,\psi]&=&\frac{1}{n!}\left(\partial_\lambda^n S_\Psi [\gb + \lambda h, \lambda \psi] \right)_{\lambda=0}\\\nonumber
&=&\frac{1}{n!}\left(\partial_\lambda^n \left(\lambda^2 S_\Psi [\gb + \lambda h,  \psi] \right)\right)_{\lambda=0}\\\nonumber
&=& \frac{1}{(n-2)!}\left(\partial_\lambda^{n-2} S_\Psi [\gb + \lambda h,  \psi]\right)_{\lambda=0}\\\nonumber
&=&\frac{1}{(n-2)!}\left[\int\!\ud^4 x h^{ab}\fd{}{\gb^{ab}}\right]^{n-2} S_\Psi[\gb,\psi].\\
\eea
Defining the partial energy momentum tensors for $h^{ab}$ and $\psi^A$  as
\bea
t^{\mathrm{g}n}_{ab} \equiv \frac{-1}{\sqrt{-\gb}}\fd{S_{\mathrm{g}n}}{\gb^{ab}}, \quad t^{\Psi n}_{ab} \equiv \frac{-1}{\sqrt{-\gb}}\fd{S_{\Psi n}}{\gb^{ab}},
\eea
respectively, we see that the partial actions are coupled as
\bea
S_n[\gb,h] = -\int\!\ud^4 x \sqrt{-\gb}h^{ab}\left(\frac{t^{\mathrm{g} n-1}_{ab}}{n}+ \frac{t^{\Psi n-1}_{ab}}{n-2}\right).
\eea
These partial actions lead to the $N^\mathrm{th}$-order equations of motion
\bea\label{matterh}
\lambda \hat{G}_{abcd} h^{cd}   &= &\kappa T^N_{ab}= \sum^N_{n=2}\lambda^n \left(t^{\mathrm{g} n}_{ab} + t^{\Psi n}_{ab} \right)\\\nonumber
\lambda \hat{W}_{AB} \psi^B  &=& \sum^N_{n=2}\bigg[ \frac{-\lambda^n}{(n-1)\sqrt{-\gb}}\\ \label{matterpsi} &&\phantom{\bigg[\sum }\times\fd{}{\psi^A}\int\!\ud^4 x \sqrt{-\gb}h^{ab}t^{\Psi n}_{ab}\bigg].
\eea
The first equation confirms that the energy-momentum tensors of $\psi^A$ and $h^{ab}$ combine as the source for the graviton. The second equation describes how the coupling between $h^{ab}$ and $t^{\Psi}_{ab}$ acts as a source for $\psi^A$. Note that even when the matter field is not free, because $S_\Psi$ never contains terms linear in the matter fields, $\hat{I}_{abA}$ must be at least linear in $\Psib^A$, so we will always have $\hat{I}_{abA}=0$ when $\Psib^A=0$.

\subsection{Non-vacuum background}
For a non-vacuum spacetime, we expect to be able to approximate (at least to first-order) the behaviour of a gravitational perturbation by ignoring the perturbations in the matter field that it might induce. Alternatively, we may have in mind a particular non-vacuum solution of the field equations $(\gb^{ab}, \Psib^A)$ and wish to find nearby solutions (approximate or exact) with precisely the same matter content. For these two scenarios, we can set $\psi^A=0$ and investigate the effect that the background $\Psib^A$ has on the dynamics of $h^{ab}$.

Considerations of this nature highlight an interesting feature of our prior discussion of the graviton action. In section \ref{act} we saw the importance of a contribution to the action $h^{ab}H_{abcd}h^{ab}$ that vanished in the vacuum; the obvious question to ask is whether a similar term exists in the non-vacuum case, and whether or not it will vanish on the \emph{non-vacuum} background equations. To answer these questions we will derive the graviton action for a non-vacuum background, which will also include the cosmological constant as a special case.

Let us restrict our attention to general relativity in the presence of a matter field:
\bea
S[g,\Psi]&=&S_\mathrm{EH}[g]+S_\Psi[g,\Psi],\\
S_\Psi[g,\Psi]&\equiv& 2\!\int\!\ud^4x\sqrt{-g}\mathcal{L}_\Psi(g^{ab}, \Psi^A, \partial_a\Psi^A).
\eea
The factor of two in the definition of the matter Lagrangian $\mathcal{L}_\Psi$ compensates for our slightly unusual normalization of $S_\mathrm{EH}$.\footnote{All our actions are twice as large as the usual definition. This normalization has no effect on the classical equations of motion, but has allowed us to define the energy-momentum tensor without a factor of two, simplifying the algebra of \S\S\ref{act}\&\ref{pert}.} It should be noted that we have assumed that $\mathcal{L}_\Psi$ does not depend on derivatives of the metric. This is the case for the Lagrangians of all the fields of the standard model except the spin-$\half$ fermion, which in any case should be coupled to gravity using the vierbein formalism, e.g.\ \cite{GTG}; such an approach is beyond the scope of this article. The results of this section can be generalized to allow $\mathcal{L}_m$ to depend on $\partial_c g^{ab}$ without any great difficulty, but this is an added algebraic complication that seems to add little insight to our investigation.

We proceed by expanding the action about a background $(\gb^{ab},\Psib^A)$ just as in (\ref{Matterexp1}) and (\ref{Matterexp2}), but now, as $\psi^A=0$, the coupling scheme (\ref{mattercouple}) reverts to the familiar energy-momentum coupling of section \ref{pert}. Following precisely the same method as section \ref{S2deriv}, we can compute $S_2$ by two successive functional derivatives (with respect to $\gb^{ab}$) applied to $S[\gb,\Psib]$. The first derivative yields
\bea
S_1 = \frac{1}{\kappa} \int\!\ud^4x \sqrt{-\gb}\left(\bar{G}_{ab}-\kappa\bar{T}^\Psi_{ab}\right)h^{ab},
\eea
where
\bea
\bar{T}^\Psi_{ab}=\frac{-1}{\sqrt{-\gb}}\fd{S_\Psi[\gb,\Psib]}{\gb^{ab}}
=-2\frac{\partial \bL_\Psi}{\partial \gb^{ab}}+ \gb_{ab} \bL_\Psi
\eea
is the energy-momentum tensor of the background matter. The second derivative yields the graviton action:
\bea\nonumber
S_2 &=& \frac{1}{2}\int\!\ud^4x h^{ab}\fd{S_1}{\gb^{ab}}\\\nonumber
&=& \frac{1}{2\kappa}\int\!\ud^4x \sqrt{-\gb} \left[h^{ab}\hat{G}_{abcd}h^{cd}\right. \\\nonumber
&&{} -\left(\bar{G}_{ab} -\kappa \bar{T}^\Psi_{ab}\right)h^{ab}h +2\kappa h^{ab}h^{cd}\frac{\partial^2\bL_\Psi}{\partial{\gb^{ab}} \partial\gb^{cd}} \\\label{S2matter}
&&{}+ \left( \bar{R} + 2\kappa\bL_\Psi\right)\left(\half h_{ab}h^{ab} - \tfrac{1}{4}h^2\right)\Big].
\eea
This is the action we sought: the generalization of equation (\ref{newact}) to a non-vacuum background. 

If we are only interested in the linear theory, and have no wish to calculate the energy-momentum tensor, then we are free to enforce the background equation 
\bea
\bar{G}_{ab} = \kappa\bar{T}^\Psi_{ab},
\eea
in the graviton action. In sharp contrast to the vacuum case, however, the background equation does not reduce $S_2$ to $\frac{1}{2\kappa}\int\!\ud^4x \sqrt{-\gb} h^{ab}\hat{G}_{abcd}h^{cd}$, or indeed any other covariantization of the massless spin-2 Fierz-Pauli action. Instead, it appears as though the background matter has endowed the graviton with mass:
\bea\label{massaction}
S_2& =& \frac{1}{2 \kappa} \int\!\ud^4x \sqrt{-\gb} \left( h^{ab}\hat{G}_{abcd}h^{cd}  + \alpha \right),
\eea
where the ``mass-term'' $\alpha$ is given by
\bea
\alpha &\equiv & - \half M \left(h^{ab}h_{ab} - \half h^2\right)+ N_{abcd}h^{ab}h^{cd},
\eea
with
\bea
M\equiv2 \kappa \left(  \bL_\Psi-\gb^{ab}\frac{\partial \bL_\Psi}{\partial \gb^{ab}} \right),\quad\!\!
N_{abcd}\equiv 2\kappa\frac{\partial^2\bL_\Psi}{\partial{\gb^{ab}} \partial\gb^{cd}}.
\eea
We refer to $\alpha$ as a ``mass-term'' because it is quadratic in $h^{ab}$, free from derivatives, and has been added to the kinetic term  $h^{ab}\hat{G}_{abcd}h^{cd}$ in the Lagrangian. However, as we will see for the specific case of the cosmological constant, $\alpha$ does not by itself determine whether the graviton is actually \emph{massive}, i.e.\ whether it propagates \emph{subluminally}; the curvature of the background will play an equally important role in the field equations. In particular, while it is tempting to identify a mass $m$ for the graviton according to $m^2=M$ (at least when $N_{abcd}=0$) we will soon see that the background matter often sets $M < 0$, so this idea is essentially untenable.

To explore these issues, it will be instructive to calculate $\alpha$ for a few simple examples. First, consider a scalar field background $\bar{\Phi}$ with Lagrangian
\bea\label{scalar}
\bL_\Phi=-\half \gb^{ab}\partial_a \bar{\Phi}\partial_b \bar{\Phi} - V(\bar{\Phi});
\eea
the mass-term is 
\bea\label{scalarmass}
\alpha_\Phi=\kappa V( \bar{\Phi}) \left( h_{ab}h^{ab} - \half h^2\right).
\eea
To ensure that the scalar field has positive energy density, we must insist that $V(\bar{\Phi})\ge 0$; hence $M \le 0$ as previously warned. Equation (\ref{scalarmass}) can also be used to find the corresponding mass-term for a cosmological constant. In this case the Lagrangian is $\mathcal{L}_\Lambda=-\Lambda/\kappa$, which we can reach from $\mathcal{L}_\Phi$ by setting $\partial_a \bar{\Phi}=0$ and $V= \Lambda/\kappa$. Clearly this gives
\bea\label{cosmass}
\alpha_\Lambda = \Lambda \left( h_{ab}h^{ab} - \half h^2\right),
\eea
which similarly suffers from $M < 0$ if the cosmological constant is positive. 

At this point, the reader may be suspicious that the formulae for $\alpha_\Phi$ and $\alpha_\Lambda$ (with $M<0$ and $N_{abcd}=0$) signify that $h^{ab}$ is a \emph{tachyon} in the presence of a scalar field background or a cosmological constant. Indeed, if the background were flat and $M$ constant over spacetime, we could derive the field equations from (\ref{massaction}), observe that their divergence enforces the de Donder gauge condition
\bea\nonumber
\partial^{\alpha}h_{\alpha\beta}- \half\partial_\beta h=0,
\eea
and, substituting this back into the equations of motion, conclude that the dynamics of the graviton were described by
\bea\nonumber
\left(\partial^2 -M\right)h^{\alpha\beta}&=&0.
\eea
This argument appears to justify the relation $m^2=M$ for the graviton's mass, and motivate the conclusion that $M < 0$ betrays tachyonic behaviour. It is important to realise, however, that the field equation above is of little relevance to the actual physical system we were discussing. In reality, $M$ will not be constant, and the presence of background matter will inevitably preclude background flatness. To understand how this last consideration alters the dynamics of the graviton, we shall briefly examine the field equation for $h^{ab}$ in the presence of a cosmological constant. First, we substitute (\ref{cosmass}) into (\ref{massaction}) and derive the field equation
\bea\label{coseom}
\hat{G}_{abcd}h^{cd} + \Lambda \left(h_{ab}- \half \gb_{ab} h \right)=0.
\eea
In contrast to the naive approach, the covariant divergence of this equation vanishes identically, and so cannot be used to relate $\delb_b h$ and $\delb^a h_{ab}$. In place of this, the gauge invariance of the vacuum theory remains intact\footnote{If we wish to extend our discussion of gauge invariance (\S\ref{GT}) to include background matter \emph{in general}, we would need to account for the gauge-fixing implicit in our starting assumption $\psi^A=0$, which is obviously not preserved by a (first-order) infinitesimal diffeomorphism $\delta \psi^A = \mathcal{L}_{\epsilon}\Psib^A$. However, because $\Lambda$ is constant over spacetime, no such difficulty arises here, and the transformations $\delta h^{ab} =-2 \delb^{(a}\epsilon^{b)}$ remain a symmetry of the equations of motion.}, and the field equation may be simplified by setting $h=0$, $\delb_a h^{ab}=0$: 
\bea\label{lambdaR}
\delb^2 h_{ab}- 2\bar{R}_{dabc}h^{dc}=0.
\eea
Surprisingly, the contribution from $\alpha_\Lambda$ has been cancelled by a term proportional to the background Ricci tensor, resulting in a field equation that is identical in form to the first-order \emph{vacuum} field equation (\ref{eqmotion1}) in this gauge. Of course, this does not indicate that the cosmological constant has no effect on the propagation of $h^{ab}$, only that these effects are limited to the constraints imposed on the background geometry by the background equation $\bar{R}_{ab}=\Lambda \gb_{ab}$. For this reason, it does not seem particularly natural to interpret $2\bar{R}_{dabc}h^{dc}$ as endowing the graviton with a mass; equation (\ref{lambdaR}) can instead be understood as a (partially gauge-fixed) massless spin-2 field equation that has been generalised to cosmological backgrounds. Quite aside from this, there is also the technical issue of interpreting the four-index tensor $\bar{R}_{abcd}$ as a mass: only if this tensor can be defined in terms of a single scalar variable (and the background metric) could the argument be made that this single variable described the graviton's mass. For a non-zero cosmological constant, the only background with this property is de Sitter space:  $\bar{R}_{dabc}= \frac{\Lambda}{3}(\gb_{db}\gb_{ac}-\gb_{dc}\gb_{ab})$, thus the gauge-fixed field equation (\ref{lambdaR}) becomes
\bea\label{desitter}
\left(\delb^2 - \frac{2\Lambda}{3}\right)h^{ab}=0.
\eea
If we were so inclined, we might interpret this as a field equation for a graviton with $m^2=2\Lambda/3$, and note that this relation has the \emph{correct} sign for positive $\Lambda$, unlike the formula $m^2=-2\Lambda$ suggested by our preliminary inspection of $\alpha_\Lambda$. In truth, however, further investigation is needed before we can either adopt or discard this interpretation. This is not only because (\ref{lambdaR}) (of which (\ref{desitter}) is a special case) can be understood as a generalisation of a massless field equation to cosmological backgrounds, but also because of the subtleties involved in interpreting the wave operator $\delb^2$ in curved space, and issues of whether or not to use a conformal coupling. Clearly, more work must be done to ascertain the physical ramifications of $\alpha_\Lambda$, and the ``mass-term'' $\alpha$ in general, before we can understand the degree to which its effects can be thought of as giving mass to the graviton.

Although massive gravitons and the cosmological constant were historically viewed as entirely separate concepts, recent work has brought to light a number of interesting connections between the two. Deser and Waldron \cite{Deserlambda} have demonstrated that, in (anti-)de Sitter background spacetimes, a massive spin-2 field is stable if and only if $m^2\ge 2\Lambda/3$, or $m=0$. While it is intriguing that our de Sitter background field equation (\ref{desitter}) suggests precisely the same special value of $m^2=2\Lambda/3$, Deser and Waldron's analysis differs significantly from our own, so this superficial observation may be misleading. In particular, whereas our mass-term  arises as a direct result of the perturbative expansion, Deser and Waldron add their mass-term to the action \emph{by hand}. Thus it is far from clear that the massive gravitons of their paper correspond to the physical system considered above. In contrast, Novello and Neves \cite{Novello} claim to prove that $m^2=-2\Lambda/3$, with the implication that $\Lambda\le0$. This approach considers an unusual generalisation of the spin-2 field equation to curved backgrounds, making a non-standard choice for the covariantization ambiguous term discussed in section \ref{FPaction}. Thus, while their calculations arguably describe a spin-2 field, this does not appear to be a natural way to describe the spin-2 field that results from perturbations of the metric (or its inverse) in Einstein's theory. It is our intention to disentangle the connections between these two approaches, and our own, in a later publication.

For the sake of completeness, we conclude this section with an example of a mass-term that can have $M>0$, and $N_{abcd}\ne0$. Unlike $\alpha_\Lambda$, however, we shall not attempt to derive any of the implications for the equations of motion. Consider an electromagnetic 1-form background $\bar{A}_a$, with Lagrangian
\bea
\bL_A= -\tfrac{1}{4}\bar{F}^2=-\tfrac{1}{4}\gb^{ab}\gb^{cd} \bar{F}_{ac}\bar{F}_{bd},
\eea
and note that $\bar{F}_{ab}\equiv 2\partial_{[a}\bar{A}_{b]}$ is independent of the metric. The calculation yields
\bea\label{elecmass}
\alpha_A = -\tfrac{1}{4}\kappa \bar{F}^2\left(h_{ab}h^{ab} - \half h^2\right) - \kappa h^{ab}h^{cd}\bar{F}_{ac}\bar{F}_{bd},
\eea
which has the aforementioned properties.

\section{Conclusion}
Contrary to the prevailing maxim, coupling the classical Fierz-Pauli graviton to its own energy and momentum \emph{does not} recreate general relativity order by order. However, there is an alternative action for the graviton (\ref{newact}) for which energy-momentum self-coupling \emph{is} consistent with Einstein's theory. Using this action, the energy-momentum tensor of the graviton (\ref{t}), added as a source to the graviton's first-order equation of motion (\ref{eqmotion1}), builds a field equation consistent with the Einstein equations to \emph{second-order}. Furthermore, the perturbative formalism developed in section \ref{pert} reveals that our action provides sufficient information to reconstruct general relativity to \emph{arbitrary} accuracy: a simple recurrence relation (\ref{emcoupling}) identifies the energy-momentum tensor at one order as the appropriate contribution to the action at the next. To any order $N$, this scheme assembles an action that dictates field equations (\ref{eom3}) in which the graviton's $N^\mathrm{th}$-order energy-momentum tensor is the source.

The formal machinery used to understand vacuum perturbations is easily extended to include matter, although the physical interpretation of the most general approach, in which matter comprises both a background field and a small perturbation, is less than transparent. Focusing on matter perturbations separately from non-vacuum backgrounds serves to clarify the formalism significantly. In a vacuum background, the interactions between the graviton and perturbations of a free matter field lead to a field equation (\ref{matterh}) in which the source for the graviton is the sum of gravitational and matter energy-momentum. This interaction inevitably induces a source in the field equations for matter (\ref{matterpsi}). Alternatively, one may neglect matter perturbations and examine the consequences of a non-vacuum background. In this case, the dynamics and energy-momentum of the graviton are prescribed by the action (\ref{S2matter}), generalizing our previous ansatz. Surprisingly, the background matter appears to induce a ``mass-term'' in the graviton action, although it is currently unclear to what extent its interpretation as a mass is valid at the level of the field equations. The mass-terms induced by a scalar field (\ref{scalarmass}), a cosmological constant (\ref{cosmass}) and electromagnetism (\ref{elecmass}) have been calculated.

\begin{acknowledgments}
L.M.B.\ is supported by STFC and St.\ John's College, Cambridge. The Mathematica package \emph{Ricci} was used for the bookkeeping part of the calculations that produced equations (\ref{t}) and (\ref{G2works}). We thank Stanley Deser and Tom\'as Ort\'in for their helpful comments, and Thanu Padmanabhan for an enlightening discussion.
\end{acknowledgments}

\appendix
\section{Padmanabhan's analysis}\label{pad}
The recent article by Padmanabhan \cite{Pad} unearths many significant shortcomings of the well known arguments \cite{krai,gupta,deser,feyngrav} that supposedly derive Einstein's equations by coupling the Fierz-Pauli graviton to its own energy-momentum tensor. Here we attempt to summarize his observations, and explain their relation to this present work.

In broad terms, Padmanabhan's criticisms fall into three areas:
\begin{enumerate}
\item The Einstein-Hilbert action consists of a bulk term (the $\Gamma^2$ action) and a surface term. The latter includes a piece \emph{linear} in $h_{\alpha \beta}$, so there can be no way to construct it from a self-coupling procedure that starts with an action that is already \emph{quadratic} in $h_{\alpha \beta}$.\footnote{The argument given by Padmanabhan is phrased in terms of non-analyticity in a dimensionful coupling constant. This form of the argument depends on his particular choice of normalization for $h_{\alpha \beta}$ and $S_\mathrm{EH}$, but is essentially equivalent to the statement given here.}
\item The starting point, the Fierz-Pauli Lagrangian (\ref{FP}), describes a \emph{Lorentz invariant} field theory, and yet the end result, general relativity, is \emph{generally covariant}. It is claimed that this metamorphosis only occurs because general covariance has been \emph{assumed} in the various derivations, in which case it is ``no big deal to obtain Einstein's theory''. More generally, the classic bootstrapping arguments wield ideas developed in general relativity (such as Hilbert's definition of the energy-momentum tensor) or use knowledge of the end result to achieve their goal. Hence they cannot be regarded as a derivation of general relativity \emph{from first principles}.
\item The first-order field equation can only take a \emph{symmetric} tensor as its source; the canonical energy-momentum tensor (\ref{Noetherdef}) is not necessarily symmetric, and although it can be made to be so, this process is not unique. Therefore the energy-momentum self-coupling procedure is ill-defined. The Hilbert definition \emph{is} uniquely determined by the action, but to use it would violate criticism 2. \emph{Crucially}, even if we allow ourselves to use Hilbert's definition, we still fail to recover the correct source term for the second-order field equation.
\end{enumerate}
It is to this very last crucial point that we have devoted the bulk of this paper. We now wish to explain our position with regards to the first two criticisms, and also Padmanabhan's proposed solution to the third.

1. Our approach expressly avoids discussing surface terms. This has greatly streamlined our formalism, and because such terms are completely irrelevant for determining field equations or energy-momentum tensors, the only price to pay for this simplicity is that we can only claim to reconstruct the Einstein-Hilbert action \emph{modulo surface terms}.\footnote{Note that this does not nessesarily mean that we have constructed the $\Gamma^2$ action, only that the integrand of the action differs from $\sqrt{-g} R$ by some total divergence.} In this sense, Padmanabhan's first criticism still stands, although it is unclear whether it has any great importance. If the action is an integral over the whole manifold, and asymptotic conditions apply to $h^{ab}$ such that the surface term at infinity vanishes, then of course there is no distinction between the Einstein-Hilbert action and the action we have constructed. Even if the action is an integral over a manifold with a boundary, so long as we consider the action to be a functional over all fields with a particular  boundary configuration (just as we might think of the action of a particle as a functional over all paths with particular end-points) the two actions differ only by an irrelevant constant. Besides, in situations where contributions from the boundary really are important, one does not typically use the Einstein-Hilbert action anyway: the Gibbons-Hawking-York boundary term \cite{Gibbons,York} must be included to remove the dependence on second derivatives of the metric. This allows the field equations to be derived using a variational principle that only demands that the variation in the fields (and not also their derivatives) vanish on the boundary.

Padmanabhan's major concern is that the surface term of the Einstein-Hilbert action has some quantum mechanical significance. As the nature of quantum gravity has yet to be understood, it remains unclear whether or not this is the case. We stress once again that the analysis in this paper is purely classical, and that we make no claims as to a quantum mechanical interpretation. Furthermore, it is not even known whether the graviton is a useful theoretical object for describing quantum gravity. We note again that the Gibbons-Hawking-York boundary term is usually included in quantum gravity investigations for which the boundary is not negligible. 

2. It is our view that Padmanabhan's concerns about general covariance are unjustified: we take the position of Weinberg \cite{Wein}, that ``general covariance by itself is empty of physical content.'' Any theory (Lorentz invariant or not) can be expressed in arbitrary curvilinear coordinates, so the requirement of general covariance cannot, in and of itself, constrain the sort of theory one might construct. Rather, the kinematical content of general relativity is encapsulated by \emph{the equivalence principle}, that the effect of gravity vanishes locally in an inertial coordinate system; thus expressing physical equations in coordinate invariant notation is an invaluable tool for describing how their dynamics are modified by gravity. It is possible that when Padmanabhan refers to `general covariance' he is referring to the equivalence principle also. As the latter is tantamount to identifying the gravitational field with a dynamical metric, he would certainly be correct to criticise any ``derivation'' that contained such a step; needless to say, we do not appeal to the equivalence principle in our approach.

General covariance aside, though, Padmanabhan's objection to the use of curved-space ideas is a valid one, indicating that none of the classic arguments constitute a derivation from first principles. Our approach certainly makes use of curved-space concepts; however our goals are perhaps not quite so bold as the other derivations that Padmanabhan has scrutinized: we do not pretend to derive general relativity purely from the ideas of Lorentz-invariant field theory. It should be stressed, however, that even if some of the \emph{kinematical} content of general relativity is in some way assumed (curved spacetime, functional derivatives with respect to the metric, etc.) it is still a ``big deal'' to derive the \emph{dynamical} content of the theory, Einstein's equations. 
 
3. We have already explained our position with regards to the definition of the energy-momentum tensor in section \ref{EMT}; the only reason that Hilbert's definition is unpalatable to Padmanabhan is that his aim is to start with as little curved-space mathematics as he can. However, the failure of the Hilbert energy-momentum tensor to give the correct second-order term for the Einstein field equations is a more significant stumbling-block. We have explained our remedy, the use of a different starting action, in the body of this paper. Padmanabhan, on the other hand, eschews energy-momentum self-coupling and introduces a new object $S^{\alpha\beta}$ that he defines with the following algorithm. Start with a Lorentz invariant Lagrangian $\mathcal{L}(\eta_{\alpha\beta},h_{\alpha\beta},\partial_\gamma h_{\alpha\beta})$ expressed in Lorentzian coordinates $\{x^\alpha\}$. Replace every instance of $\eta_{\alpha\beta}$ with the metric $\gb_{\alpha\beta}$ to produce a new Lagrangian $\widetilde{\mathcal{L}}(\gb_{\alpha\beta},h_{\alpha\beta},\partial_\gamma h_{\alpha\beta})$; note that this is \emph{not} the same as expressing $\mathcal{L}$ in an arbitrary coordinate system because the partial derivatives $\partial_\alpha$ have not been upgraded to covariant derivatives $\delb_\alpha$. We can now define
\bea\label{Sdef}
S^{\alpha\beta}\equiv 2\left. \frac{\partial \sqrt{-\gb} \widetilde{\mathcal{L}}}{\partial\gb_{\alpha\beta}}\right|_{\gb=\eta}.
\eea
The subscript reminds us that we must set $\gb_{\alpha\beta}=\eta_{\alpha\beta}$ after taking the metric derivative, as we are supposedly working in Lorentzian coordinates. Padmanabhan claims to be able to reconstruct the $\Gamma^2$ action by coupling $h_{\alpha\beta}$ to this new object $S^{\alpha\beta}$. Unfortunately $S^{\alpha\beta}$ has a number of highly undesirable properties, suggesting that it is a rather unnatural object, ill-defined in its current form.\footnote{In private communication, Padmanabhan has indicated that he shares our concerns about $S^{\alpha\beta}$ and does not believe it to be of any fundamental importance; hence we present the case against $S^{\alpha\beta}$ for the sake of completeness rather than rebuttal.}

Firstly, as it has been constructed from a Lagrangian rather than an action, $S^{\alpha\beta}$ depends directly on surface terms. This introduces a very large ambiguity, as $S^{\alpha\beta}$ will depend on whether we write the integrand of the action in the form $(\partial h)^2$, as Padmanabhan does, in the form $h\partial^2h$, or as some arbitrary combination of both. Each possibility defines a different $S^{\alpha\beta}$ and (presumably) leads to a different self-coupled limit for the graviton. It seems that the only remedy for this ambiguity is to artificially stipulate that $\mathcal{L}$ contain no second derivatives, although we note in passing that even this leaves us free to add surface terms of the form $\partial^\alpha(\phi A_\alpha)$ in theories for fields other than the graviton.

The second troubling aspect to $S^{\alpha\beta}$ is the ``half-covariantizing'' algorithm used to construct $\widetilde{\mathcal{L}}$. It should be clear that this procedure has only been defined in Lorentzian coordinates, thus the matrix $S^{\alpha\beta}$ does not really constitute the components of a tensor, as we have not explained how their values change when expressed in another coordinate system.\footnote{The insistence that we be able to calculate the components of this object in arbitrary coordinates has nothing to do with curved spacetime or general relativity. Rather, this reflects the perfectly reasonable expectation that we should be able to express Padmanabhan's self-coupling procedure in \emph{flat-space} spherical polar coordinates, for example, or any other coordinate system we choose.} There are essentially two ways to extend the definition (\ref{Sdef}) to include curvilinear coordinates. The trivial solution is to construct the tensor $S^{ab}\equiv S^{\alpha\beta}(\partial_\alpha)^a(\partial_\beta)^b$ using the vectors $\{(\partial_\alpha)^a\}$, partial derivatives with respect to the Lorentzian coordinates used to calculate $S^{\alpha\beta}$ in the first place. This obviously defines a genuine tensor, so the components $S^{\alpha^\prime \beta^\prime }$ of $S^{ab}$ in some curvilinear coordinate system $\{ x^{\alpha^\prime} \}$ can be calculated, and they will be related to $S^{\alpha\beta}$ by the usual transformation rules. It should be clear, however, that this solution is rather unnatural: suppose we have a Lagrangian expressed in a curvilinear coordinate system, then the only way to calculate the components $S^{\alpha^\prime\beta^\prime}$ in that system is to first transform to Lorentzian coordinates, calculate $S^{\alpha\beta}$ according to (\ref{Sdef}), and then transform back to our original coordinate system. Also, because this process picks out a special set of coordinates, there is also no reason to expect that $S^{ab}$ can be written as a tensorial function of $h_{ab}$, $\gb_{ab}$ and $\delb_a$. The \emph{natural} way to proceed would be to generalize the definition (\ref{Sdef}) in such a way that we could calculate $S^{\alpha^\prime\beta^\prime}$ working in any coordinate system. It might seem that a viable solution would be to define the tensor
\bea\label{Sdef2}
S^{ab}\equiv \frac{2}{\sqrt{-\gb}} \left. \frac{\partial \sqrt{-\gb} \mathcal{L}} {\partial\gb_{ab}}\right|_{\bar{\Gamma}},
\eea
where $\mathcal{L}=\mathcal{L}(\gb_{ab},h_{ab},\delb_{c} h_{ab})$ is the \emph{fully} covariant Lagrangian, and the subscript  indicates that the Christoffel symbols $\bar{\Gamma}^{a}_{\phantom{a}bc}$ are to be treated as independent of the metric and held constant in the derivative. This expression generalizes (\ref{Sdef}) to define a tensor $S^{ab}$ in a coordinate invariant fashion; because the Christoffel symbols are held constant, no term arises from a variation of the covariant derivatives, and $S^{ab}$ will reduce to $S^{\alpha\beta}$ in Lorentzian coordinates. This expression gives us some insight into the geometrical meaning of Padmanabhan's half-covariantized algorithm; in particular it reveals that the derivative $\partial/\partial\gb_{\alpha\beta}$ used to define $S^{\alpha\beta}$ is in fact exploring geometries (infinitesimally close to Minkowski spacetime) with connections that are not metric compatible.\footnote{This is the same operation as the derivative used to acquire the Einstein equations from the Palatini action \cite{palant}, although here we will have no cause to perform the complementary derivative $\partial/\partial \Gamma|_{\gb}$.} It is perhaps unsurprising that this $\bar{\Gamma}$-constant derivative introduces a new layer of ambiguity to the procedure, as we can now alter $S^{ab}$ by adding terms proportional to $0=\delb_c \gb_{ab}$ to the Lagrangian. Although this might seem a rather contrived objection, it is in fact a very common consideration. For example, suppose the Lagrangian includes a term of the form $\delb_a h^a_{\phantom{a}b}$; should we calculate $S^{ab}$ by acting with $ \partial/\partial \gb |_{\bar{\Gamma}}$ on $\delb_a (\gb^{ac} h_{cb})$, or should we first commute the metric past the covariant derivative, and act on $\gb^{ac}\delb_a  h_{cb}$ instead? Note that this issue would have been invisible in Lorentzian coordinates because
\bea
\left.\frac{\partial \delb_c \gb_{ef}} {\partial\gb_{ab}}\right|_{\bar{\Gamma}}= - 2 \bar{\Gamma}^{(a}_{\phantom{(a}c(e} \delta^{b)}_{f)},
\eea
which we would have automatically set to zero. It seems the only way to avoid this uncertainty in $S^{ab}$ is to introduce another artificial constraint on the Lagrangian: we insist that it be written in such a way that no derivatives act on the metric. This should be achieved by commuting covariant derivatives through the metric, rather than  integrating by parts, due to the aforementioned issues with surface terms.

We shall take our analysis of $S^{\alpha\beta}$ no further at this time. It is still uncertain whether this object can be generalized, naturally and uniquely, to form a genuine tensor; without such a generalization it is difficult to ascertain what sort of mathematical object the matrix of functions $S^{\alpha\beta}$ is supposed to represent. Although we cannot claim to have exhausted all possibilities, the evidence before us suggests, at the very least, that this goal is not easily achieved.

Aside from these technical issues, we should also emphasize that, unlike the energy-momentum tensor, $S^{\alpha\beta}$ has no apparent physical interpretation beyond its supposed role in a graviton self-coupling scheme. Energy-momentum self-coupling was justified by analogy with matter-gravity coupling, and advanced by the notion that the energy-momentum of \emph{all} fields should source gravitation. In contrast, the self-coupling scheme involving $S^{\alpha\beta}$ only serves to set gravity apart from the other fields. Furthermore, our solution displays an unusual symmetry between the coupling terms in the action and source terms generated in the field equations as a result (see \S\ref{EMtensor}); this symmetry is broken by Padmanabhan's self-coupling procedure.

\section{Expansion of $G_{ab}$}\label{calc}
Here we determine the first two terms of the expansion of the Einstein tensor
\bea
G_{ab}= G^{(1)}_{ab} + G^{(2)}_{ab} + O(h^3),
\eea
induced by a perturbation of the inverse metric about a vacuum background:
\bea
g^{ab}&=& \gb^{ab}+ h^{ab},\\
\bar{G}_{ab} &=&0.
\eea
The perturbation in the metric is of course fixed by the relationship $g^{ab}g_{bc}=\delta^a_c$,
\bea
\Rightarrow \quad g_{ab}= \gb_{ab} - h_{ab} + h_{ac}h^c_{\phantom{c}b} + O(h^3).
\eea
To begin, introduce a connection $E^{a}_{\phantom{a}bc}$ between the derivative operators $\nabla_a$ and $\delb_a$: 
\bea\label{Edef}
E^{a}_{\phantom{a}bc} = \half g^{ab}(\delb_b g_{cd}+\delb_c g_{bd}-\delb_d g_{bc}).
\eea
This allow the Ricci tensor to be expressed as
\bea
R_{ab}=2\left(\delb_{[c}E^{c}_{\phantom{c}a]b} + E^{c}_{\phantom{c}d[c}E^{d}_{\phantom{d}a]b}\right).
\eea
From (\ref{Edef}) it is clear that
\bea
E^{a(0)}_{\phantom{a}bc}&=&0,\\\label{E1}
E^{a(1)}_{\phantom{a}bc}&=&-\half \gb^{ad}(2\delb_{(b} h_{c)d}-\delb_d h_{bc}),\\\nonumber
E^{a(2)}_{\phantom{a}bc}&=&-\half h^{ad}(2\delb_{(b} h_{c)d}-\delb_d h_{bc})\\\label{E2}
&&{}+\half \gb^{ad}(2\delb_{(b} (h_{c)e}h^e_{\phantom{e}d})-\delb_d (h_{be}h^e_{\phantom{e}c})).
\eea
Hence the terms of the expansion $R_{ab}=R^{(1)}_{ab} + R^{(2)}_{ab} + O(h^3)$ can be computed as follows:
\bea\label{R1}
R^{(1)}_{ab}&=& 2\delb_{[c}E^{c(1)}_{\phantom{c}a]b}\\\label{R2}
R^{(2)}_{ab}&=&2\left(\delb_{[c}E^{c(2)}_{\phantom{c}a]b} + E^{c(1)}_{\phantom{c}d[c}E^{d(1)}_{\phantom{d}a]b}\right).
\eea
Thus,
\bea\nonumber
G^{(1)}_{ab}&=& R^{(1)}_{ab} - \half \gb_{ab}R^{(1)}_{cd}\gb^{cd}\\\nonumber
&=& -\delb_c \delb_{(a}h_{b)}^{\phantom{a)}c} + \half \delb^2 h_{ab} + \half\delb_a\delb_b h\\
&&{}- \half\gb_{ab}\left( - \delb_c\delb_d h^{cd} + \delb^2 h \right) ,
\eea
which confirms that $\hat{G}_{abcd}$, as defined in (\ref{Gdef}), represents the linearised Einstein tensor:
\bea\label{G(1)}
\hat{G}_{abcd}h^{cd} = G^{(1)}_{ab}.
\eea
In particular, note that both sides of this equation agree on the order of the derivatives in $\delb_c \delb_{(a}h_{b)}^{\phantom{a)}c}$; this is the descendant of the covariantization ambiguous term discussed in section \ref{FPaction}.

To find $G^{(2)}_{ab}$, start with 
\bea\nonumber
G^{(2)}_{ab}&=& R^{(2)}_{ab} - \half \gb_{ab}\left(R^{(2)}_{cd}\gb^{cd} + R^{(1)}_{cd}h^{cd}\right)\\&&{}+ \half h_{ab}R^{(1)}_{cd}\gb^{cd},
\eea
and substitute equations (\ref{R1}) and (\ref{R2}), followed by (\ref{E1}) and (\ref{E2}). The bookkeeping for this calculation is characteristically laborious, but is easily accomplished using a computer algebra package; the result is
\bea\label{G2works}
G_{ab}^{(2)}=-\kappa t_{ab} + \half h\hat{G}_{abcd} h^{cd},
\eea
where $t_{ab}$ is given by (\ref{t}). As expounded in section \ref{fieldeq}, and now confirmed by direct calculation (\ref{G(1)}), the first-order approximation to the Einstein field equation is $\hat{G}_{abcd} h^{cd}=0$, so $\hat{G}_{abcd} h^{cd} = O(h^2)$ must hold true at second-order. Clearly it follows from this that $h\hat{G}_{abcd} h^{cd} = O(h^3)$, and hence (\ref{works}) is verified.

The third-order difference between $G_{ab}^{(2)}$ and $-\kappa t_{ab}$ exists because the field equation approximated to second-order in (\ref{EOM2}) is actually $\sqrt{-g}G^{ab}/\sqrt{-\gb}=0$; this is of course entirely equivalent to the usual form of the Einstein field equation $G_{ab}=0$.

\bibliography{act}

\begin{thebibliography}{20}
\expandafter\ifx\csname natexlab\endcsname\relax\def\natexlab#1{#1}\fi
\expandafter\ifx\csname bibnamefont\endcsname\relax
  \def\bibnamefont#1{#1}\fi
\expandafter\ifx\csname bibfnamefont\endcsname\relax
  \def\bibfnamefont#1{#1}\fi
\expandafter\ifx\csname citenamefont\endcsname\relax
  \def\citenamefont#1{#1}\fi
\expandafter\ifx\csname url\endcsname\relax
  \def\url#1{\texttt{#1}}\fi
\expandafter\ifx\csname urlprefix\endcsname\relax\def\urlprefix{URL }\fi
\providecommand{\bibinfo}[2]{#2}
\providecommand{\eprint}[2][]{\url{#2}}

\bibitem[{\citenamefont{Padmanabhan}(2008)}]{Pad}
\bibinfo{author}{\bibfnamefont{T.}~\bibnamefont{Padmanabhan}},
  \bibinfo{journal}{Int.J.Mod.Phys. D} \textbf{\bibinfo{volume}{17}},
  \bibinfo{pages}{367} (\bibinfo{year}{2008}),
  \urlprefix\url{arXiv:gr-qc/0409089v1}.

\bibitem[{\citenamefont{Deser}(1970)}]{deser}
\bibinfo{author}{\bibfnamefont{S.}~\bibnamefont{Deser}}, \bibinfo{journal}{Gen.
  Rel. Grav.} \textbf{\bibinfo{volume}{1}}, \bibinfo{pages}{9}
  (\bibinfo{year}{1970}).

\bibitem[{\citenamefont{Feynman et~al.}(1995)\citenamefont{Feynman, Morinigo,
  and Wagner}}]{feyngrav}
\bibinfo{author}{\bibfnamefont{R.~P.} \bibnamefont{Feynman}},
  \bibinfo{author}{\bibfnamefont{F.~B.} \bibnamefont{Morinigo}},
  \bibnamefont{and} \bibinfo{author}{\bibfnamefont{W.~G.}
  \bibnamefont{Wagner}}, \emph{\bibinfo{title}{Feynman Lectures on
  Gravitation}} (\bibinfo{publisher}{Addison-Wesley}, \bibinfo{year}{1995}),
  pp. \bibinfo{pages}{74--88}.

\bibitem[{\citenamefont{Gupta}(1954)}]{gupta}
\bibinfo{author}{\bibfnamefont{S.~N.} \bibnamefont{Gupta}},
  \bibinfo{journal}{Phys. Rev.} \textbf{\bibinfo{volume}{96}},
  \bibinfo{pages}{1683} (\bibinfo{year}{1954}).

\bibitem[{\citenamefont{Kraichnan}(1955)}]{krai}
\bibinfo{author}{\bibfnamefont{R.~H.} \bibnamefont{Kraichnan}},
  \bibinfo{journal}{Phys. Rev.} \textbf{\bibinfo{volume}{98}},
  \bibinfo{pages}{1118} (\bibinfo{year}{1955}).

\bibitem[{\citenamefont{Deser}(1987)}]{Deser2}
\bibinfo{author}{\bibfnamefont{S.}~\bibnamefont{Deser}},
  \bibinfo{journal}{Classical and Quantum Gravity}
  \textbf{\bibinfo{volume}{4}}, \bibinfo{pages}{L99} (\bibinfo{year}{1987}),
  \urlprefix\url{http://stacks.iop.org/0264-9381/4/L99}.

\bibitem[{\citenamefont{Boulware and Deser}(1975)}]{Deser&Boulware}
\bibinfo{author}{\bibfnamefont{D.~G.} \bibnamefont{Boulware}} \bibnamefont{and}
  \bibinfo{author}{\bibfnamefont{S.}~\bibnamefont{Deser}},
  \bibinfo{journal}{Annals of Physics} \textbf{\bibinfo{volume}{89}},
  \bibinfo{pages}{193 } (\bibinfo{year}{1975}), ISSN \bibinfo{issn}{0003-4916}.

\bibitem[{\citenamefont{Ortin}(2004)}]{Ortin}
\bibinfo{author}{\bibfnamefont{T.}~\bibnamefont{Ortin}},
  \emph{\bibinfo{title}{Gravity and Strings}} (\bibinfo{publisher}{Cambridge
  University Press}, \bibinfo{year}{2004}), chap. \bibinfo{chapter}{3.2}.

\bibitem[{\citenamefont{Fierz and Pauli}(1939)}]{fierz}
\bibinfo{author}{\bibfnamefont{M.}~\bibnamefont{Fierz}} \bibnamefont{and}
  \bibinfo{author}{\bibfnamefont{W.}~\bibnamefont{Pauli}},
  \bibinfo{journal}{Proceedings of the Royal Society of London. Series A,
  Mathematical and Physical Sciences} \textbf{\bibinfo{volume}{173}},
  \bibinfo{pages}{211} (\bibinfo{year}{1939}), ISSN \bibinfo{issn}{00804630},
  \urlprefix\url{http://www.jstor.org/stable/97457}.

\bibitem[{\citenamefont{Wald}(1984)}]{wald}
\bibinfo{author}{\bibfnamefont{R.~M.} \bibnamefont{Wald}},
  \emph{\bibinfo{title}{General Relativity}} (\bibinfo{publisher}{University of
  Chicago Press}, \bibinfo{year}{1984}), p. \bibinfo{pages}{437}.

\bibitem[{\citenamefont{Kuchar}(1976)}]{kuch}
\bibinfo{author}{\bibfnamefont{K.}~\bibnamefont{Kuchar}},
  \bibinfo{journal}{Journal of Mathematical Physics}
  \textbf{\bibinfo{volume}{17}}, \bibinfo{pages}{801} (\bibinfo{year}{1976}),
  \urlprefix\url{http://link.aip.org/link/?JMP/17/801/1}.

\bibitem[{\citenamefont{Mannheim}(2006)}]{man}
\bibinfo{author}{\bibfnamefont{P.~D.} \bibnamefont{Mannheim}},
  \bibinfo{journal}{Physical Review D (Particles, Fields, Gravitation, and
  Cosmology)} \textbf{\bibinfo{volume}{74}}, \bibinfo{eid}{024019}
  (pages~\bibinfo{numpages}{6}) (\bibinfo{year}{2006}),
  \urlprefix\url{http://link.aps.org/abstract/PRD/v74/e024019}.

\bibitem[{\citenamefont{Magnano and Sokolowski}(2002)}]{Mag}
\bibinfo{author}{\bibfnamefont{G.}~\bibnamefont{Magnano}} \bibnamefont{and}
  \bibinfo{author}{\bibfnamefont{L.~M.} \bibnamefont{Sokolowski}},
  \bibinfo{journal}{Classical and Quantum Gravity}
  \textbf{\bibinfo{volume}{19}}, \bibinfo{pages}{223} (\bibinfo{year}{2002}),
  \urlprefix\url{http://stacks.iop.org/0264-9381/19/223}.

\bibitem[{\citenamefont{Lasenby et~al.}(1998)\citenamefont{Lasenby, Doran, and
  Gull}}]{GTG}
\bibinfo{author}{\bibfnamefont{A.}~\bibnamefont{Lasenby}},
  \bibinfo{author}{\bibfnamefont{C.}~\bibnamefont{Doran}}, \bibnamefont{and}
  \bibinfo{author}{\bibfnamefont{S.}~\bibnamefont{Gull}},
  \bibinfo{journal}{Phil. Trans. R. Soc. Lond. A}
  \textbf{\bibinfo{volume}{356}}, \bibinfo{pages}{487} (\bibinfo{year}{1998}),
  \urlprefix\url{arXiv:gr-qc/0405033v1}.

\bibitem[{\citenamefont{Deser and Waldron}(2001)}]{Deserlambda}
\bibinfo{author}{\bibfnamefont{S.}~\bibnamefont{Deser}} \bibnamefont{and}
  \bibinfo{author}{\bibfnamefont{A.}~\bibnamefont{Waldron}},
  \bibinfo{journal}{Physics Letters B} \textbf{\bibinfo{volume}{508}},
  \bibinfo{pages}{347 } (\bibinfo{year}{2001}), ISSN \bibinfo{issn}{0370-2693}.

\bibitem[{\citenamefont{Novello and Neves}(2003)}]{Novello}
\bibinfo{author}{\bibfnamefont{M.}~\bibnamefont{Novello}} \bibnamefont{and}
  \bibinfo{author}{\bibfnamefont{R.~P.} \bibnamefont{Neves}},
  \bibinfo{journal}{Classical and Quantum Gravity}
  \textbf{\bibinfo{volume}{20}}, \bibinfo{pages}{L67} (\bibinfo{year}{2003}),
  \urlprefix\url{http://stacks.iop.org/0264-9381/20/L67}.

\bibitem[{\citenamefont{York}(1972)}]{York}
\bibinfo{author}{\bibfnamefont{J.~W.} \bibnamefont{York}},
  \bibinfo{journal}{Phys. Rev. Lett.} \textbf{\bibinfo{volume}{28}},
  \bibinfo{pages}{1082} (\bibinfo{year}{1972}).

\bibitem[{\citenamefont{Gibbons and Hawking}(1977)}]{Gibbons}
\bibinfo{author}{\bibfnamefont{G.~W.} \bibnamefont{Gibbons}} \bibnamefont{and}
  \bibinfo{author}{\bibfnamefont{S.~W.} \bibnamefont{Hawking}},
  \bibinfo{journal}{Phys. Rev. D} \textbf{\bibinfo{volume}{15}},
  \bibinfo{pages}{2752} (\bibinfo{year}{1977}).

\bibitem[{\citenamefont{Weinberg}(1972)}]{Wein}
\bibinfo{author}{\bibfnamefont{S.}~\bibnamefont{Weinberg}},
  \emph{\bibinfo{title}{Gravitation and Cosmology}}
  (\bibinfo{publisher}{Wiley}, \bibinfo{year}{1972}), pp.
  \bibinfo{pages}{91--93}.

\bibitem[{\citenamefont{Hobson et~al.}(2006)\citenamefont{Hobson, Efstathiou,
  and Lasenby}}]{palant}
\bibinfo{author}{\bibfnamefont{M.~P.} \bibnamefont{Hobson}},
  \bibinfo{author}{\bibfnamefont{G.~P.} \bibnamefont{Efstathiou}},
  \bibnamefont{and} \bibinfo{author}{\bibfnamefont{A.~N.}
  \bibnamefont{Lasenby}}, \emph{\bibinfo{title}{General Relativity: An
  Introduction for Physicists}} (\bibinfo{publisher}{Cambridge University
  Press}, \bibinfo{year}{2006}), chap. \bibinfo{chapter}{19.10}.

\end{thebibliography}
\end{document}